\documentclass[final,3p,times]{elsarticle}

\usepackage[OT1]{fontenc}
\usepackage{amsmath,mathtools}
\usepackage{amssymb}
\usepackage{upquote}
\usepackage{cadabra} 
\usepackage[colorlinks]{hyperref}
\usepackage{dirtree}
\biboptions{compress}

\def\ContinueLineNumber{\lstset{firstnumber=last}} 
\def\StartLineAt#1{\lstset{firstnumber=#1}}

\newcommand{\p}[2]{{}^{^{(#1)}}\! #2}

\newcommand{\boxedFunction}[2]{\vspace*{3mm}
\noindent
\fbox{\begin{minipage}{46em}
\texttt{#1}\\\emph{\footnotesize #2}
\end{minipage}}\newline 
}

\journal{Computer Physics Communications}

\begin{document}

\begin{frontmatter}
  \title{Cadabra and Python algorithms in General Relativity and Cosmology II: Gravitational Waves}
  \author[1]{Oscar Castillo-Felisola}
  \ead{o.castillo.felisola@protonmail.com}


  \author[2]{Dominic T. Price}
  \ead{dominicprice@outlook.com}

  \author[3]{Mattia Scomparin\corref{cor1}}
  \ead{mattia.scompa@gmail.com}
  \cortext[cor1]{Corresponding author}

  \address[1]{Departamento de F\'isica and Centro Cient\'ifico y
    Tecnol\'ogico de Valpara\'iso (CCTVal),\\Universidad T\'ecnica Federico
    Santa Mar\'ia\\Casilla 110-V, Valpara\'iso, Chile}

  \address[2]{Department of Mathematical Sciences, Durham
    University,\\South Road, Durham DH1 3LE, United Kingdom}

  \address[3]{Via del Grano 33,\\
    Mogliano Veneto, Italy}

  \begin{abstract}
    Computer Algebra Systems (CASs) like \verb+Cadabra+ Software play a
    prominent role in a wide range of research activities in physics and
    related fields. We show how \verb+Cadabra+ language is easily
    implemented in the well established Python programming framework,
    gaining excellent flexibility and customization to address the issue
    of tensor perturbations in General Relativity. We obtain a
    performing algorithm to decompose tensorial quantities up to any
    perturbative order of the metric. The features of our code are
    tested by discussing some concrete computational issues in research
    activities related to first/higher-order gravitational waves.
  \end{abstract}

  \begin{keyword}
    Computer Algebra System, \verb+Cadabra+, Gravitation, Classical Field
    theory, Gravitational Waves, Perturbative Field Theory, Python,
    Cosmology, General Relativity. 
  \end{keyword}
\end{frontmatter}


\section{Introduction}
\label{sec:intro}

This article is the continuation of our previous work
\cite{castillo-felisola20_cadab_python_algor_gener_relat_cosmol_i}
describing the use of \verb+Cadabra+ algorithms within the framework
of General Relativity and Cosmology. \verb+Cadabra+ is an open-source
Computer Algebra System (CAS) designed for field-theory problems and
specialized for both abstract and component computations. The \verb+Cadabra+
language is Python-oriented and works closely with the Python library for
symbolic calculations \emph{SymPy}, but uses its own Python class for
storing expressions and uses \LaTeX{} formatting for inputs and outputs.

In the last few years many introductory works about \verb+Cadabra+
\cite{peeters07_cadab,peeters07_introd_cadab,peeters07_symbol_field_theor_with_cadab,Brewin:2019qbs}
have appeared. Among all the applications described in this wide bibliography
include problems dealing with field theory, differential equations
and symbolic matrix algebra. Nevertheless, only a small number of
them discuss the currently available spectrum of physical applications
which development of \verb+Cadabra+ algorithms supported by
Python coding allows. Thanks to its many language constructs and wide 
ranging standard library, Python is a very
complete and flexible programming language employed in many
high-performance scientific back-end and front-end applications. In
addition, Python offers a wide range of external libraries such as
\emph{NumPy}, \emph{SymPy}, \emph{SciPy} and \emph{Matplotlib} which
make it a powerful and up-to-date ecosystem for dealing with mathematical,
statistical, biological, human and social sciences. From this
perspective, an in-depth investigation into how \verb+Cadabra+ algorithms
could be consolidated and enhanced with Python is strongly
motivated.

Of course there are many other CASs available and which are used in 
research on a daily basis and which contain specialist packages and functionality
for dealing with a wide range of problems, however one of the main motivations
for the writing of this paper is to display how even without dedicated functionality
already available in the language it is not difficult to create very powerful
and generic programs by using Python with \verb+Cadabra+. In particular, we wish to draw
to the attention of the reader the following constructs which differ from other
CASs and which are used throughout this paper when constructing algorithms
and performing other manipulations:
\begin{itemize}
\item Expressions in \verb+Cadabra+ are mutable which provides powerful ways to interact
with them, especially by combining this with Python iterator constructions which
allows expressions to be visited and modified using \verb|for| loops and other
intuitive constructs.
\item By allowing expressions to be queried for different property types, `smart'
algorithms can be written which respond differently to different types of input.
Properties can also be dynamically attached to new symbols which allows the
construction of functions which programmatically define new sets of objects.
\item Python has many useful and performant inbuilt containers which have
rich interfaces, and also provides implementations of many more useful container
types such as \verb|defaultdict| in the \verb|collections| library. These make
storing and accessing related expressions very easy and natural which improves the
organisation of the code.
\item As \verb+Cadabra+ uses a \LaTeX{}format for inputting and outputting expressions, 
expressions can be created or manipulated by using Python's string and regex
functions if a feature is not implemented in \verb+Cadabra+. While this is not always
an optimal solution, it is one of the great strengths of Python which makes it
such a productive language that it is almost always possible to achieve some end.
\end{itemize}

In order to showcase as many of these features as possible, we have chosen to
take a more in-depth look at tensor perturbation theory in General Relativity
which is by no means the only field topic whose analysis is assisted by using
the tools Python and \verb+Cadabra+ offer; but as it is a very well studied topic with a
lot of literature dedicated to it we hope this makes the paper accessible to a large
audience. Tensor perturbations, also known as Gravitational Waves,
are disturbances in the curvature of the spacetime metric and represent
one of the most definitive theoretical and phenomenological signatures
of General Relativity as a standard theory of gravitation. Recently,
gravitational waves have attracted a wide interest thanks to the
detections made by the LIGO and Virgo interferometers sourced by compact
binary coalescences
\cite{Abbott:2016blz,Abbott:2016nmj,Abbott:2017vtc,TheLIGOScientific:2017qsa,GBM:2017lvd,Monitor:2017mdv,Abbott:2020khf}.

Having taken gravitational waves as the most effective case study to examine, the
main purpose of this work is therefore to put forth the first structured
insight on the approach, design, and implementation of hybrid
\verb+Cadabra+ and Python functions/commands and methods of defining a new coding
environment within \verb+Cadabra+. In particular, we use our
multi-annual experience gained in both \verb+Cadabra+ and Python
to provide a programming vision of the entire life-cycle of such a
development process, which consists of five phases as displayed in
Table \ref{tab:phases}.
\begin{table}[!h]
  \centering
  \begin{tabular}{||l|l|l||}
    \hline
    Step & Description & Ref. Secs.\\
    \hline\hline
    Theory & Analysis, modelling, formalism, reference objects & Sec. \ref{sec:tensPertGR}\\ 
    \hline
    Ref. libraries development & Supporting functions \& environments, coding core paradigms& Sec. \ref{sec:lib}\\ 
    \hline
    Main algorithms development& Code development, optimization, new user-oriented commands& Sec. \ref{sec:tenspert}\\ 
    \hline
    Notebooks& Explore the new environment, upgrade \verb+Cadabra+, communicate& Sec. \ref{sec:fistgw}, \ref{sec:gwrel}, \ref{sec:highgw} \\ 
    \hline
    Testing \& Discussion& Timing, performance, perspectives \& generalizations & Sec. \ref{sec:concl}\\ 
    \hline
  \end{tabular}
  \caption{Steps of the development process with related reference sections of this article.}
  \label{tab:phases}
\end{table}

When talking about gravitational waves, methods related to
\textit{Perturbation Theory} are essential and Section
\ref{sec:tensPertGR} is dedicated to providing a general overview of
tensorial perturbative expansions in General Relativity. In
particular, the spacetime metric representing the gravitational field
is treated as a series of successive, increasingly small tensor
perturbations around a background, where an $N$-th order gravitational
wave solution is obtained by (i) truncating the series by keeping only
the first $N$ terms, and (ii) forcing consistency with perturbed
General Relativity equations and gauge conditions.

Operating tensorial expansions within \verb+Cadabra+ naturally leads to us
introducing in Section \ref{sec:lib} a set of new hybrid \verb+Cadabra+
and Python functions (see Table \ref{tab:pertliblist}) collected
inside a library which we call \verb+perturbations.cnb+. These functions
are united by the goal of defining the basic perturbative elements
we will deal with. They fix the core assumptions of our algorithms
including the formalism, e.g. how perturbations will be represented,
and the new user-oriented interface commands, e.g. how to invoke
the new perturbative objects obtained in the code. This step is deeply
influenced by the Pythonic aesthetic in favour of writing functions that are
perfect for the immediate use-case, and which can be developed and
tested separately before finally being put together.

Section \ref{sec:tenspert} presents our first result, an algorithm
called \verb+perturb()+ able to deal with tensor perturbations for
every order of a generic tensor defined in terms of the spacetime
metric. The following example illustrates the use of our machinery to
perturb the Cristoffel connection $\Gamma^{\mu}\,_{\nu \tau}$, stored
in the \verb+ch+ variable, up to the second order of metric tensor:
\begin{cadabra}[numbers=none]
connection = perturb(ch,[gLow,gUpp],'pert',2)
\end{cadabra}
where \verb+gLow+ and \verb+gUpp+ hold information about the metric tensor.

%
%

The code returns the Python dictionary \verb+connection+ which
contains useful representations of the perturbation such as a symbolic
decomposition in the \verb|'sym'| key: 
\begin{dgroup*}
  \begin{dmath*}
    \Gamma^{\mu}\,_{\nu \tau} = \p{\Gamma}^{\mu}\,_{\nu \tau} +
    \p{1}{\Gamma}^{\mu}\,_{\nu \tau} + \p{2}{\Gamma}^{\mu}\,_{\nu
      \tau}
  \end{dmath*}
  \begin{dsuspend}
    and an array of the perturbative orders in the \verb|'ord'| key,
    so in this case \verb|connection['ord'][2]| is
  \end{dsuspend}
  \begin{dmath*}
    \p{2}{\Gamma}^{\mu}\,_{\nu \tau} =
    \tfrac{1}{2}\partial_{\tau}{\p{2}{h}_{\nu}\,^{\mu}}+\tfrac{1}{2}\partial_{\nu}{\p{2}{h}_{\tau}\,^{\mu}}
    - \tfrac{1}{2}\partial^{\mu}{\p{2}{h}_{\nu \tau}} -
    \tfrac{1}{2}\p{1}{h}^{\mu \sigma} \partial_{\tau}{\p{1}{h}_{\nu
        \sigma}} - \tfrac{1}{2}\p{1}{h}^{\mu \sigma}
    \partial_{\nu}{\p{1}{h}_{\tau \sigma}}+\tfrac{1}{2}\p{1}{h}^{\mu
      \sigma} \partial_{\sigma}{\p{1}{h}_{\nu \tau}}
  \end{dmath*}
\end{dgroup*}
More information about the data structure can be found in Table \ref{tab:sometab}.


The algorithm is very performative and has been developed to be both
flexible and intuitive, goals which the use of the timing analysis
tools introduced in our fist paper
\cite{castillo-felisola20_cadab_python_algor_gener_relat_cosmol_i} has
helped us to achieve. The results of this analysis are reported in
Section \ref{sec:concl}. The  \verb+perturb()+ function takes full
consideration of the symmetries and other fundamental properties of
the perturbed object. This section also presents several examples of
computations, focusing on the dependence of timings on properties of
perturbed expressions. 

The suite composed by the \verb+perturbations.cnb+ library and the
\verb+perturb()+ algorithm defines a new \verb+Cadabra+ perturbative
environment that must be explored and used alongside normal
\verb+Cadabra+ notebooks. As toy models to provide a staging ground,
Section \ref{sec:fistgw} and Section \ref{sec:gwrel} are completely
dedicated to first order gravitational waves, obtained in the
so-called harmonic gauge, with the respective sections calculating (i)
the wave-equation which describes the propagation of first order
gravitational waves, and (ii) the related wave-relations. In the last
Section \ref{sec:highgw} we apply our machinery to reproduce some
results obtained in Ref. \cite{Arcos:2015uqa}, where a complete
analytical analysis of vacuum high-order gravitational waves solutions
is given. Hence, we provide a computational counterpart of such
analysis, highlighting the natural predisposition of \verb+Cadabra+
Software and the newly developed perturbative environment to treat the
heavy and onerous nature of high-order tensor calculations. In
particular, we discuss how to obtain the numerical coefficients of
higher-order gravitational wave solutions. 

A final discussion is drawn in Section \ref{sec:concl} where we
analyse the timing, performance, and perspectives of our new
perturbative environment, highlighting the full consistency of our
results with the predicted exponential timing behaviour derived from
the number of perturbed terms.

The experience gained through all phases of the development process,
from designing to release, took us into a leading role in the design
of \verb+Cadabra+ and Python algorithms. During the development
process we needed to identify the algorithms' bottlenecks and boost
their performance to satisfy notebooks' demands about practical
manipulating, debugging, and programming issues not covered by
\verb+Cadabra+ itself.

With reference to the process followed, we believe that there are some
important takeaways of programming in \verb+Cadabra+:
\begin{itemize}
\item  \textit{Developing Cadabra algorithms supported by Python offers superior algorithm solutions.} The well-established coding vision coming from Python suggests a more structured way of thinking about algorithms and provides a powerful factory of tools that can inspire many improvements of \verb+Cadabra+.
\item \textit{Cadabra architecture is completely predisposed to work with Python data-structures}. \verb+Cadabra+ expression are one of the main paradigms of \verb+Cadabra+'s architecture and they are Python objects that can be treated as data inside more complex algorithm structures to be processed.
\item \textit{Cadabra algorithms can be debugged and tested in a very smart way}. We exhibit in the first part \cite{castillo-felisola20_cadab_python_algor_gener_relat_cosmol_i} for the first time some new approaches on \verb+Cadabra+ code debugging. A new approach to test timing performance has been introduced within the new \verb+timing.cnb+ library.
\item \textit{The development of new computing environments induces an
    improvement of Cadabra's core}. When working with our new
  perturbative environment, we needed to introduce new functions into
  \verb+Cadabra+'s core library, exhibiting a wide range of possible
  improvements for \verb+Cadabra+. The fact that the introduction of a new
  environment has induced an improvement of the standard \verb+Cadabra+
  environment shows that \verb+Cadabra+ has new undiscovered domains of
  improvements that approaches like ours can discover. From this
  perspective, \verb+Cadabra+'s code and community mutually support and
  inspire progress with each other. 
\end{itemize}
Finally, it is important to notice that the results developed in this
paper can be easily customized with respect to any physical model to
which a perturbative approach must be applied, as is the case with
statistical mechanics \cite{THOULESS1960553}, radiative transfer
\cite{BOX200295}, quantum mechanics \cite{Picasso2009}, particle
physics \cite{Cvetic:2011vz}, fluid mechanics \cite{1975STIA},
chemistry \cite{Ayers}, and many other contexts. Our algorithms are
built in such a way that one can cut-and-paste expressions straight
from this paper into a \verb+Cadabra+ notebook and customize the
content according to specific needs. The main notebooks of this work
will be completely available on the official \verb+GitLab+
repository.\footnote{See~\url{https://gitlab.com/cdbgr/cadabra-gravity-II}.}

Since new functions have been introduced in the \verb+Cadabra+ core
libraries, our code requires a minimum of version \verb+2.3.8+ of
\verb+Cadabra+ software, and as it is heavily based on the cooperation
of \verb+Cadabra+ with Python it is completely incompatible with the
old \verb+1.x+ versions.

\section{Overview of Tensor Perturbations in General Relativity}
\label{sec:tensPertGR}

There is a vast, practically limitless, choice of works aimed to
introducing the basics of General Relativity and its implications in
Cosmology (see for example
Refs.~\cite{Weinberg:1972kfs,Carroll:2004st}). In our analysis, we
will follow the notations and conventions defined in the first part of
our previous
work~\cite{castillo-felisola20_cadab_python_algor_gener_relat_cosmol_i}. 

As a starting point, let us consider spacetimes affected by small
perturbations $h_{\mu\nu}$ about Minkowski spacetime $\eta_{\mu\nu}$.
More precisely, the metric tensor $g_{\mu\nu}$ can be written 
\begin{equation}
  g_{\mu\nu} \equiv \p{0}{g}_{\mu\nu} + h_{\mu\nu} 
  = \eta_{\mu\nu} + h_{\mu\nu} \simeq \eta_{\mu\nu} + \p{1}{h}_{\mu\nu}.
\end{equation}
Note that the linear approximation has been employed and higher order
terms than first have been neglected. Consequently, as
$g_{\alpha\mu}g^{\mu\beta}=\delta_\alpha^\beta$, the contravariant
metric can be written as $g^{\mu\nu} = \eta^{\mu\nu} - h^{\mu\nu}$,
with $h^{\mu\nu}=\eta^{\mu\alpha}\eta^{\nu\beta}h_{\alpha\beta}$. As a
consequence, it is clear that the indices of all first-order tensorial
quantities will be raised or lowered using the flat Minkowski metric.
Adopting such perturbative decompositions, it is simple to exhibit
that the first order Einstein's equations 
\begin{equation}
  \label{eq:einsimple}
  \p{1}{R_{\mu\nu}}  = \kappa \p{1}{S}_{\mu \nu},
\end{equation}
can be rewritten as  
\begin{equation}
  \label{eq:ff}
  \tfrac{1}{2}\partial_{\nu \rho}{\p{1}{h}_{\mu}\,^{\rho}}
  - \tfrac{1}{2}\partial^{\rho}\,_{\rho}{\p{1}{h}_{\mu \nu}}
  - \tfrac{1}{2}\partial_{\mu \nu}{\p{1}{h}_{\rho}\,^{\rho}}
  + \tfrac{1}{2}\partial_{\mu}\,^{\rho}{\p{1}{h}_{\nu \rho}}
  = \kappa \p{1}{S}_{\mu \nu},
\end{equation}
with $\kappa$ the constant gravitational coupling and
$\p{1}{S}_{\mu \nu}$ the first-order term of the source tensor, \(S\),
defined as
\begin{equation*}
  S_{\mu \nu} = T_{\mu \nu} - \frac{1}{2} g_{\mu \nu} T.
\end{equation*}

This equation does not possess unique solutions since, given a
solution, it will always be possible to identify one another solution
performing a coordinate transformation of the form 
\begin{equation}
  x^{\prime\alpha} = x^\alpha + \xi^{\alpha}(x^\alpha),
\end{equation}
where $\partial_{\beta} \xi^{\alpha}$ is of the same order of
$h_{\mu\nu}$. This property is known as \emph{gauge invariance}. The
redundancy can be removed by fixing a specific coordinate system. For
our purposes, it is a good choice to work in the so-called
\emph{harmonic gauge}, defined by the condition  
\begin{equation}
  \label{eq:xcvxx}
  0 = \Gamma^\sigma = g^{\mu \nu} \Gamma^{\sigma}{}_{\mu \nu},
\end{equation}
which perturbed up to first order and inserted in Eq. \eqref{eq:ff} in
vacuum yields the first-order gravitational wave equation 
\begin{equation}\label{eq:vcvx}
  \partial^{\alpha} \partial_{\alpha} \p{1}{h}_{\mu\nu} = 0.
\end{equation}
Without loss of generality, Eq. \eqref{eq:vcvx} is solved by tensor
plane-wave parametrization solutions that can be parametrized as
follows: 
\begin{equation}\label{eq:plane}
  \p{1}{h}_{\mu \nu}
  =
  \mathbf{e}_{\mu \nu} \exp\left(i k_{\lambda} x^{\lambda}\right)
  + \bar{\mathbf{e}}_{\mu \nu} \exp\left(-i k_{\lambda} x^{\lambda}\right),
\end{equation}
with 
\begin{equation}\label{eq:wrelll}
  k_{\lambda} k^{\lambda} = 0
  \quad
  \text{and}
  \quad
  \mathbf{e}^{\lambda}\,_{\nu} k_{\lambda} - \tfrac{1}{2}\mathbf{e}^{\mu}\,_{\mu}  k_{\nu} = 0.
\end{equation}
Above, we have introduced the spacetime coordinate $x^\mu$, the
wave-number $k_\lambda$ and the symmetric polarization tensor
$\mathbf{e}_{\mu\nu}$ with its complex conjugate
$\bar{\mathbf{e}}_{\mu\nu}$. Eqs. \eqref{eq:wrelll} are commonly known
as the \emph{first-order wave relations}. 

The entire framework described in the present section can be extended
to higher-order perturbations. As we will see in the following
sections, the covariant metric can be decomposed as 
\begin{equation}
  g_{\mu\nu}
  \equiv
  \sum_{n=0}^{\bar{n}} \p{n}{g}_{\mu\nu}
  =
  \sum_{n=0}^{\bar{n}} f_n \left( \eta, \p{1}{h}, \cdots, \p{n}{h} \right).
\end{equation}
In our notation, $\p{n}{Y}$ stands for the $n$-th perturbative order
associated to the tensor $Y$, with ${\bar{n}}$ the higher perturbative
order we are interested in.
 
In such higher-order context the methods described for first-order
perturbations and concept of gauge invariance hold. With reference to
the main equations, Eqs. \eqref{eq:xcvxx} and \eqref{eq:vcvx} become
\begin{equation}
  \label{eq:pertseq}
  0 = \p{n}{\Gamma}^{\sigma}\left( \eta, \p{1}{h}, \cdots, \p{n}{h} \right)
  \text{ and }
  0 = \p{n}{R}_{\mu\nu}\left( \eta, \p{1}{h}, \cdots, \p{n}{h} \right)
  \text{ with }
  n \in \{0,...,\bar{n}\}.
\end{equation}
In particular, wave solutions propagating in vacuum satisfying Eqs.
\eqref{eq:pertseq} admit a very general parametrization  
\begin{equation}
  \label{eq:parsolggh}
  \p{2n}{h_{\mu\nu}}
  =
  \Phi^n \left\{ \p{2n}{a} \, \eta_{\mu\sigma} \alpha^{\sigma}{}_{\nu} 
    +
    i z \omega c^{-1}
    \p{2n}{b} \, \eta_{\mu\sigma}\beta^{\sigma}{}_{\nu} \right\}
  \exp \left(i 2 n k_{\rho}x^{\rho}\right) + c.c.,
\end{equation} 
with, among all the objects introduced,
$\Phi \equiv {\mathbf{e}}^{\mu}\,_{\nu} {\mathbf{e}}^{\nu}\,_{\mu}$
and $\alpha^{\sigma}{}_{\nu}$, $\beta^{\sigma}{}_{\nu}$ constant
matrices. The numerical coefficients
\begin{equation}
  \left\{\p{2n}{a}, \p{2n}{b}\,\right\} \text{ for } n = 1, 2, \ldots,
\end{equation}
characterize each perturbative order solution. Parametrization
\eqref{eq:parsolggh} means that the odd-order solutions do not
contribute to the metric perturbation since one can see that all
odd-order terms in the metric perturbation vanish identically. This
feature may be related to the physical properties of the gravitational
interaction \cite{Arcos:2015uqa}.

\section{Libraries}
\label{sec:lib}

\begin{figure}[h!!!]
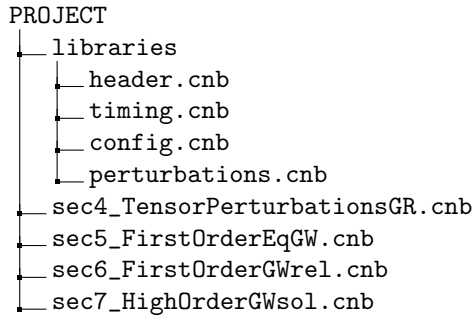
 
  \centering
\begin{minipage}{7cm}
\dirtree{%
.1 PROJECT.
.2 libraries.
.3 header.cnb.
.3 timing.cnb.
.3 config.cnb.
.3 perturbations.cnb.
.2 sec4_TensorPerturbationsGR.cnb.
.2 sec5_FirstOrderEqGW.cnb.
.2 sec6_FirstOrderGWrel.cnb.
.2 sec7_HighOrderGWsol.cnb.
 }
\end{minipage}
\caption{Rooting of the coding.}
\label{fig:root}
\end{figure}

The code which accompanies this paper is organised into a number of
modules, each of which contains functions and computations
corresponding to a different section. See Figure \ref{fig:root} for the rooting. 
This section will introduce the
content in the \emph{libraries} directory, which contains general
routines which are used throughout the remainder of the paper: 
(i) \emph{header.cnb} setting up a common environment of property and 
object definitions, (ii) \emph{perturbations.cnb} which defines general
perturbative algorithms which are the building blocks of the
discussion.

The remaining files in the \emph{algorithms} directory are the
notebooks which accompany the other sections in this paper;
\verb+sec4_TensorPerturbationsGR.cnb+ defines an algorithm able to
deal with tensor perturbations up to every order of a generic tensor
defined in terms of the metric tensor (section \ref{sec:tenspert}),
\verb+sec5_FirstOrderEqGW.cnb+ to get first-order gravitational waves
equation (section \ref{sec:fistgw}), \verb+sec6_FirstOrderGWrel.cnb+
to get first-order gravitational wave relations (section
\ref{sec:gwrel}), and \verb+sec7_HighOrderGWsol.cnb+ to study
higher-order gravitational wave solutions (section \ref{sec:highgw}).

\subsection{The header.cnb library}
\label{sec:head}
The header library is the same introduced in
\cite{castillo-felisola20_cadab_python_algor_gener_relat_cosmol_i}. We
refer to such paper for more information.

\subsection{The config.cnb library: Defining global configuration variables}
\label{sec:config}

This package contains global configuration variables accessible from
the notebooks. Its content simply defines the perturbation label and
the maximum order in perturbation,
\begin{cadabra}
maxPertOrder = 2
pertLabel = 'pert'
\end{cadabra}

It is particularly useful for setting the perturbation order of
notebooks which differ from the default value by setting it before
importing other notebooks, e.g. using
\begin{cadabra}[numbers=none]
import libraries.config as config
config.maxPertOrder = 6

from sec4_TensorPerturbations import * 
\end{cadabra}
will now use \verb+maxPertOrder = 6+, overriding the default value.

\subsection{The perturbations.cnb library}
\label{sec:func1}

This package contains various helper functions to define a standard approach to formally perturb expressions and equations in elementary ways, e.g. 
formally define standard perturbative formalism for \verb+Cadabra+, inherit properties from unperturbed objects to perturbed ones, 
give perturbative decompositions of expressions and equations, substitute perturbative expansions. In order to use this package, we will import it with the standard
Python import statement \verb+import libraries.perturbations as perturb+. 

Uploading the \verb+perturbations+ library, the following suite of functions will be available,
of which we summarize the perspective of purpose in Table \ref{tab:pertliblist}.

\begin{table}[!h]
  \centering
  \begin{tabular}{||l|c|c||}
    \hline
    Function & Input example& Output example\\
    \hline\hline
    \verb+defPertSymbol+ & $n,A$ & $\p{n}{A}$\\ 
    \hline
    \verb+defPertList+ & $\bar{n},A$ & $\p{0}{A},\ldots,\p{\bar{n}}{A}$\\ 
    \hline
    \verb+defPertSum+ & $\bar{n},A$ & $A=\p{0}{A}+\ldots+\p{\bar{n}}{A}$\\ 
    \hline
             & $A=B$ & \\ 
    \verb+subsPertSums+ & $A=\p{0}{A}+\ldots+\p{\bar{n}}{A}$ & $\p{0}{A}+...+\p{\bar{n}}{A}=\p{0}{B}+...+\p{\bar{n}}{B}$\\ 
             & $B=\p{0}{B}+\ldots+\p{\bar{n}}{B}$ & \\ 
    \hline
    \verb+getEquationPertOrder+ & $n,\p{0}{A}+...+\p{\bar{n}}{A}=\p{0}{B}+...+\p{\bar{n}}{B}$ & $\p{n}{A}=\p{n}{B}$\\ 
    \hline
  \end{tabular}
  \caption{The suite of functions defined in \texttt{perturbations}
    library with some examples. The $\p{n}{A}$ and $\p{n}{B}$ symbols denote
    perturbations, with associated weights and properties. The $\bar{n}$
    variable stands for the higher perturbative order of the
    decompositions. In particular, $n \le \bar{n}$.} 
  \label{tab:pertliblist}
\end{table}

This subsection is dedicated to the definition of appropriate
functions aimed at customizing the standard \verb+Cadabra+ commands with
Python, adapting them to our purposes. Although the purpose is not
immediately understood, the usefulness will be evident in the
following 

First of all we  define a test environment, namely objects and
properties to test our functions. 
\begin{cadabra}[numbers=none]
# Definitions for test cases
\partial{#}::PartialDerivative.
M{#}::LaTeXForm("\hat{\Psi}").
R{#}::LaTeXForm("\hat{\Phi}").
{M_{\mu\nu},R_{\mu\nu}}::Symmetric.
{T{#},M{#}}::Depends(\partial{#}).
\end{cadabra}
\boxedFunction{defPertSymbol(ex:Ex, pertLabel:str, pertOrder:int) -> Ex}
{Returns a new object with the same structure as \texttt{ex} but with
  \texttt{pertOrder} appended to its name, and assigns the properties
  of \texttt{ex} to this new object} 

A really useful function to introduce is \verb+defPertSymbol+, that,
given a \verb+Cadabra+ object, a perturbative label, and a
perturbative  order, returns the associated the symbolic-perturbed
Cadabra object. Formally defines standard perturbative formalism for
Cadabra,  inheriting properties from unperturbed objects to perturbed
ones and fixing a perturbation label. Note: the function inherits only
\verb|Symmetric|, \verb|TableauSymmetry| and \verb|Depends|
properties: other needed properties  can be added by customizing the
last block code at the end of the function, where such proprieties are
queried to the \verb|Ex| object.
\StartLineAt{2}
\begin{cadabra}
def defPertSymbol (ex, pertLabel, pertOrder): 
  # Create the perturbed version of the object by appending pertOrder to its name
  pert = ex.copy()
  pert.top().name += str(pertOrder)
  # Create anoter version where the indices are replaced by {#}
  pertGeneral = pert.copy()
  for index in pertGeneral.top().indices():
    index.erase()
  pertGeneral.top().append_child(${#}$)
  # Get the LaTeX form of ex with no indices
  exName = ex.copy()
  for index in exName.top().indices():
    index.erase()
  exLaTeX = exName._latex_()
  # Copy the LaTeX form of ex to pertGeneral, with the pertOrder written prefix and superscript
  LaTeXForm(pertGeneral, Ex(r'"\,^{^{(
  # Assign properties of ex to pert and pertGeneral
  Weight(pertGeneral, Ex(f'label={pertLabel}, value={pertOrder}'))
  p_sym = Symmetric.get(ex)
  if p_sym is not None:
    p_sym.attach(pert)
  p_depends = Depends.get(ex)
  if p_depends is not None:
    p_depends.attach(pertGeneral)
  p_tab = TableauSymmetry.get(ex)
  if p_tab is not None:
    p_tab.attach(pert)
  return pert
\end{cadabra}

Note that the function automatically recognizes some fundamental
properties of the cadabra unperturbed object, and inheriths them
inside the perturbed symbol. Additionally, let us highlight the use of
tableaux symmetry. Some tests
\begin{cadabra}[numbers=none]
defPertSymbol($M_{\mu\nu}$,'pert',7);
\end{cadabra}
\begin{dmath*}
  \p{7}{\hat{\Psi}_{\mu \nu}}
\end{dmath*}
that takes the original \verb+Cadabra+ object $\hat{\Psi}_{\mu \nu}$ and
gives the associated seventh-order perturbed symbol
$\p{7}{\hat{\Psi}_{\mu \nu}}$. In particular, the following tests
hold: 
\begin{cadabra}[numbers=none]
# Testing Inherited Symmetry
x1 = canonicalise($M7_{\mu\nu}+M7_{\nu\mu}$);
# Testing Dependencies
x2 = unwrap($a \partial_{\rho}{M7_{\mu\nu}}$);
# Testing Inherited Weight
x3 = keep_weight($M7_{\mu\nu}$,$pert=7$);
x4 = keep_weight($M7_{\mu\nu}$,$pert=5$);
x5 = keep_weight($M7_{\mu}^{\nu}$,$pert=7$);
x6 = keep_weight($M7_{\mu}^{\nu}$,$pert=5$);
\end{cadabra}
as can be easily seen in the Table \ref{tab:root}:
\begin{table}[h]
  \centering
  \begin{tabular}{||l|l|c|c|c||}
    \hline
    Ex& Tested Property & Input & Output & Passed \\
    \hline\hline
    \texttt{x1} & \texttt{Symmetry} & $\p{7}{\hat{\Psi}_{\mu \nu}}+\p{7}{\hat{\Psi}_{\nu \mu}}$ & $2\p{7}{\hat{\Psi}_{\mu \nu}}$ & $\checkmark$\\ 
    \hline
    \texttt{x2} & \texttt{Dependencies}  &$ \partial_{\rho}{(a\p{7}{\hat{\Psi}_{\mu \nu}})}$ & $a \partial_{\rho}{\p{7}{\hat{\Psi}_{\mu \nu}}}$ & $\checkmark$ \\
    \hline
    \texttt{x3} & \texttt{Weight} & $\p{7}{\hat{\Psi}}_{\mu \nu}$& $\p{7}{\hat{\Psi}_{\mu \nu}}$ & $\checkmark$ \\ 
    \hline
    \texttt{x4} & \texttt{Weight} & $\p{7}{\hat{\Psi}}_{\mu \nu}$& $0$ & $\checkmark$ \\ 
    \hline
    \texttt{x5} & \texttt{Weight} & $\p{7}{\hat{\Psi}}_{\mu}\,^{\nu}$& $\p{7}{\hat{\Psi}_{\mu}}\,^{\nu}$ & $\checkmark$ \\ 
    \hline
    \texttt{x6} & \texttt{Weight} & $\p{7}{\hat{\Psi}}_{\mu}\,^{\nu}$& $0$ & $\checkmark$ \\ 
    \hline
  \end{tabular}
  \caption{Test results for the \texttt{defPertSymbol}  function.}
  \label{tab:root}
\end{table}

Note that after the use of \verb|defPertSymbol($a_{\mu\nu}$,'x',2)|,
the new symbol \verb|y:=a2_{\mu\nu}| results defined as
$\p{2}{a_{\mu\nu}}$. Also note that such function, once used, fixes
weights and dependencies independentely from type, position and number
of indices, i.e. the previous example defines also the perturbative
quantities (and properties) for $\p{2}{a_{\mu\nu\rho}}$, and
$\p{2}{a_{\mu}^{\nu}}$, and the other mixtures of indices.

Another function that we will use frequently is
\texttt{defPertList()}.

\boxedFunction{defPertList(ex:Ex, pertLabel:str, maxPertOrder:int) -> List[Ex]}
{Calls \texttt{defPertSymbol} on \texttt{ex} for all values of
  \texttt{pertOrder} up to and including \texttt{maxPertOrder}, and
  returns a list of all generated objects}

It could be extremely useful for our scope
define a set of perturbed symbols. This is exactly the purpose of the
getPertList function, which takes a \texttt{Cadabra} object and
produces with respect to it a Python list containing for each position
from 0 (background) to \texttt{maxPertOrder} (the higher perturbative
order desired) the associated perturbative symbol generated by
\texttt{defPertSymbol()}. Referring to the passed arguments, such
symbol will be characterized by a weight proportional to the array
position and labelled by \texttt{pertLabel}. Formally the function
defines all the perturbative symbols of an object from background to a
max perturbative order.
\StartLineAt{30}
\begin{cadabra}
def defPertList (ex, pertLabel, maxPertOrder):
  pertList = []
  for i in range(maxPertOrder+1):
    symbol_i = defPertSymbol(ex,pertLabel,i) 
    pertList.append(symbol_i)
  return pertList 
\end{cadabra}
We thought it might be useful to highlight the initialization of the
Python array \texttt{pertLs=[]} and the use of the
\texttt{pertLs.append()} to valorize each position of the array with a
Cadabra object in succession. By way of example, the following command 
\begin{cadabra}[numbers=none]
x = defPertList($M_{\mu\nu}$,'pert',2)
x[1];
\end{cadabra}
\begin{dmath*}
  \p{1}{\hat{\Psi}_{\mu \nu}}
\end{dmath*}
generates the list $[\p{0}{\hat{\Psi}_{\mu \nu}},\p{1}{\hat{\Psi}_{\mu
    \nu}},\p{2}{\hat{\Psi}_{\mu \nu}}]$, where the $n$-th element has
weight equal to $n$ and inherits the the same dependencies of the
$M_{\mu\nu}$ operator. 

\boxedFunction{defPertSum(ex:Ex, exPertList:List[Ex]) -> Ex} {Return
  an equation with \texttt{ex} on the left hand side, and a sum of the
  terms in \texttt{exPertList} on the right hand side i.e. write
  \texttt{ex} as a sum of its perturbative components} 
  
Formally defines the decomposition of an \verb+Ex+ object as the sum of
perturbative orders contained in \verb+exPertList+.

\StartLineAt{36}
\begin{cadabra}
def defPertSum(ex, exPertList):
  pertSum := 0:
  for x in exPertList : pertSum += x
  return $@(ex) = @(pertSum)$
\end{cadabra}
As an example, consider the following case
\begin{cadabra}[numbers=none]
y1 := M_{\mu}:
y2 = defPertList(y1,'pert',3)
y3 = defPertSum(y1,y2);
\end{cadabra}
\begin{dmath*}
  \hat{\Psi}_{\mu} = \p{0}{\hat{\Psi}_{\mu}}+\p{1}{\hat{\Psi}_{\mu}}+\p{2}{\hat{\Psi}_{\mu}}+\p{3}{\hat{\Psi}_{\mu}}
\end{dmath*}

\boxedFunction{subsPertSums(ex:Ex, partLabel:str, maxPertOrder:int, *pertSums:Ex) -> Ex}
{Substitute \texttt{pertSums} into \texttt{ex}, distribute the result
  and drop any terms higher than \texttt{maxPertOrder}}

Substitutes a set of \texttt{Ex pertSums} into a given \texttt{Ex}
object. The function automatically drops all terms higher than a
\texttt{maxPertOrder} with respect to a certain \texttt{partLabel}.
The function automatically distinguishes if the expression is or not
an equation (drop weight never works fine in equations, hence we need
to separate the terms and then rebuild the equation). \StartLineAt{40}
\begin{cadabra}
def subsPertSums(ex, pertLabel, maxPertOrder, *pertSums):
  # Substitute and distribute over the result
  for pertSum in pertSums: 
    substitute(ex, pertSum)
  distribute(ex)
  # Remove higher order terms
  if ex.top().name == r"\equals":
    # Equation, need to drop left and right hand side separately
    l, r = lhs(ex), rhs(ex)
    for i in range(maxPertOrder+1,2*maxPertOrder+1):
      drop_weight(l, Ex(f"{pertLabel}={i}"))
      drop_weight(r, Ex(f"{pertLabel}={i}"))
    return $@(l) = @(r)$
  else:
    for i in range(maxPertOrder+1,2*maxPertOrder+1):
      drop_weight(ex, Ex(f"{pertLabel}={i}"))
    return ex
\end{cadabra}
The Python \verb+*+ symbol allows to pass an arbitrary number of
argument objects to the function and entrusts a good flexibility about
the function. The validity of the substitutions lies in the correct
declarations passed through \texttt{*pertSums}.  

The effective functioning of this function can be seen with the
following example: 
\begin{cadabra}[numbers=none]
z1 := R_{\mu}:
z2 = defPertList(z1,'pert',3)
z3 = defPertSum(z1,z2)
w = subsPertSums($M_{\mu}R_{\nu}$,'pert',3,y3,z3);
\end{cadabra}
This code replaces inside the expression \verb+M_{\mu}R_{\nu}+ the
\verb+defPertSum+ of \verb+M_{\mu}+ and \verb+R_{\nu}+ separately: 
\begin{dmath*}
  \p{0}{\hat{\Psi}_{\mu}} \p{0}{\hat{\Phi}_{\nu}}
  +
  \p{0}{\hat{\Psi}_{\mu}} \p{1}{\hat{\Phi}_{\nu}}
  +
  \p{0}{\hat{\Psi}_{\mu}} \p{2}{\hat{\Phi}_{\nu}}
  +
  \p{0}{\hat{\Psi}_{\mu}} \p{3}{\hat{\Phi}_{\nu}}
  +
  \p{1}{\hat{\Psi}_{\mu}} \p{0}{\hat{\Phi}_{\nu}}
  +
  \p{1}{\hat{\Psi}_{\mu}} \p{1}{\hat{\Phi}_{\nu}}
  +
  \p{1}{\hat{\Psi}_{\mu}} \p{2}{\hat{\Phi}_{\nu}}
  +
  \p{2}{\hat{\Psi}_{\mu}} \p{0}{\hat{\Phi}_{\nu}}
  +
  \p{2}{\hat{\Psi}_{\mu}} \p{1}{\hat{\Phi}_{\nu}}
  +
  \p{3}{\hat{\Psi}_{\mu}} \p{0}{\hat{\Phi}_{\nu}}
\end{dmath*}
Note that terms with an higher order than \texttt{maxPertOrder} equal
to $3$ are removed! Very useful for long computations to avoid terms
that will be neglected! 

\boxedFunction{getEquationPertOrder(ex:Ex, pertLabel:str, pertOrder:int) -> Ex}
{Return \texttt{ex} where all terms of an order different to
  \texttt{pertOrder} are discarded}

The function \verb+getEquationPertOrder()+ gets the specified
perturbative order from an equation. We introduce this function
because the standard command keep_weight doesn't work with equations
but only with expressions. Given an expanded equation, gets an
estabilished perturbative order
\StartLineAt{57}
\begin{cadabra}
def getEquationPertOrder (ex, pertLabel, pertOrder):
  l = keep_weight(lhs(ex), Ex(f"{pertLabel}={pertOrder}"))
  r = keep_weight(rhs(ex), Ex(f"{pertLabel}={pertOrder}"))
  return $@(l) = @(r)$
\end{cadabra}
As an example, the action of the function is as follows
\begin{cadabra}[numbers=none]
m = subsPertSums($M_{\mu}=R_{\mu}$,'pert',3,y3,z3);
getEquationPertOrder(m,'pert',2);
\end{cadabra}
\begin{dgroup*}[noalign]
  \begin{dmath*}
    \p{0}{\hat{\Psi}_{\mu}}+\p{1}{\hat{\Psi}_{\mu}}+\p{2}{\hat{\Psi}_{\mu}}+\p{3}{\hat{\Psi}_{\mu}} = \p{0}{\hat{\Phi}_{\mu}}+\p{1}{\hat{\Phi}_{\mu}}+\p{2}{\hat{\Phi}_{\mu}}+\p{3}{\hat{\Phi}_{\mu}}
  \end{dmath*}
  \begin{dmath*}
    \p{2}{\hat{\Psi}_{\mu}} = \p{2}{\hat{\Phi}_{\mu}}
  \end{dmath*}
\end{dgroup*}
where only second order terms are extracted from the initial equation.

\section{Tensor perturbations in General Relativity}
\label{sec:tenspert}

The following section aims to introduce and implement a set of useful
algorithms designed to make \texttt{Cadabra} a powerful environment to
deal with the fundamental symbolical computations related to
gravitational waves field. In particular, our goal is to introduce a
standard package of functions within \verb+Cadabra+ software to deal
with tensor perturbations up to every \texttt{maxPertOrder} order of
every metric tensor object.

In particular, in this notebook we derive the covariant expansion of
the metric tensor in terms of tensor perturbations. We extend this
result to fundamental tensor objects in General Relativity (such as
the Riemann tensor, Ricci tensor and others). 

\subsection{Setting the perturbative environment}
\label{sec:env}

For this example we shall import the \verb+header+, \verb+config+ and
\verb+perturbation+ modules provided with the bundle of notebooks.
Other standard \texttt{Cadabra} modules are imported by the
\verb+header+ itself.
\StartLineAt{1}
\begin{cadabra}
from libraries.header import *
from libraries.perturbations import *
from libraries.config import maxPertOrder, pertLabel
\end{cadabra}
See Section \ref{sec:lib} for more information.

To characterize our perturbative calculations must define  the higher
perturbative order we will deal with. Hence we finish setting up the
common environment by calling \verb+init_properties+ and defining the
highest order to which we will be performing our calculations.
\ContinueLineNumber
\begin{cadabra}
init_properties(coordinates=$t,x,y,z$, metrics=[$g_{\mu\nu}$, $\eta_{\mu\nu}$])
\end{cadabra}

We finish the initialisation by assigning weights to the objects that
will appear in our perturbative expansion 
\begin{cadabra}
\eta{#}::Weight(label=pert,type=multiplicative,value=0).
\delta{#}::WeightInherit(label=pert,type=multiplicative,value=1).
\partial{#}::WeightInherit(label=pert,type=multiplicative,value=1).
\end{cadabra}
As can be easily seen, such operators are weighted as background
quantities. 

\subsection{Covariant Metric Perturbations \textnormal{\texttt{gLow[]}}}
\label{sec:mlow}

Let us consider a gravitational wave sourced at great distances from
the Earth, e.g by two black holes mergers or two neuron star mergers.
Due to the distance, the amplitude of the gravitational wave will be
very small when it arrives on Earth and this assumption justifies us
to put the following decomposition of the  covariant metric
tensor: 
\begin{equation}\label{eq:pertCovg}
g_{\mu\nu}=\eta_{\mu\nu}+\p{1}{h_{\mu\nu}}+\p{2}{h_{\mu\nu}}\,,
\end{equation}
where $\eta_{\mu\nu}$ is the flat covariant metric tensor and
$\p{i}{h_{\mu\nu}}$ are small successive higher order tensor
perturbations. In our notation, \texttt{maxPertOrder} stands for the
maximum perturbative order with witch we want to deal with.
Computationally, definition \eqref{eq:pertCovg} can be implemented in
a very smart way. First of all, considering the perturbative unit
$h_{\mu\nu}$, its related properties can be defined as follows: 
\begin{cadabra}
tensorPert:=h_{\mu\nu}:
# Properties
{h_{\mu\nu},h^{\mu\nu}}::Symmetric.
h{#}::Depends(\partial{#}).
\end{cadabra}
To begin, let us obtain  a Python list of \verb+Cadabra+ perturbation
symbols $[\p{0}{g_{\mu\nu}},\p{1}{g_{\mu\nu}},\p{2}{g_{\mu\nu}}]$ by
using our \verb+defPertList+ function 
\begin{cadabra}
gPertList = defPertList($g_{\mu\nu}$,pertLabel,maxPertOrder)
\end{cadabra}
The defPertList function defines, without showing them, the objects
\texttt{g0}, \texttt{g1} \texttt{g2} that we can recall every time we
need. Finally notice that the \verb+defPertList+ function assigns to
\verb+gLow['ord']+ a Python list, so no initialization
\verb+gLow['ord']=[]+ as an empty list is needed. 

Let us also initialize an empty Python dictionary:
\begin{cadabra}
gLow = {} 
\end{cadabra}
A dictionary is a collection of objects or variables which can be
saved inside it unordered, changeable and indexed through a key item.
In Python dictionaries, you can access the items of a dictionary by
referring to its key name.  For the sake of clarity on our approach,
in this work each dictionary will be the key unit containing all the
information for each perturbed tensor. As a matter of fact, we will
also have a specific dictionary for the connection, a specific
dictionary for the Ricci tensor and so on. All these dictionaries will
have in common, a part trivial cases, the key names with which access
to their values. Key names have a specific meaning referring to the
perturbative information we want to use.

On that note, our model of dictionary is desribed in Table
\ref{tab:sometab}, where we use the \texttt{gLow} as a prototype, the
same rules will be extended to all the other dictionaries

\begin{table}[!h]
  \centering
  \begin{tabular}{|| l l l l ||} 
    \hline
    Key& Code & Output & Description \\
    \hline\hline
    \texttt{unp} & \verb+gLow['unp']+ & $g_{\mu\nu}$ & Unperturbed expression\\ 
    \hline
    \texttt{ord} & \verb+gLow['ord'][1]+  & $\p{1}{h_{\mu\nu}}$ & Perturbative orders of the expression into a list \\
    \hline
    \texttt{sym} & \verb+gLow['sym']+ & $g_{\mu\nu}=\p{0}{g_{\mu\nu}}+\p{1}{g_{\mu\nu}}+\p{2}{g_{\mu\nu}}$ & Symbolic decomposition of the expression \\ 
    \hline
  \end{tabular}
  \caption{Model key names and the specific perturbative information we
    want to use in our dictionary structure.}
  \label{tab:sometab}
\end{table}

Anyway, focusing back up for \texttt{gLow}, to valorize the dictionary
structure described in Table \ref{tab:sometab}, we can use the
following code 
\begin{cadabra}
gLow['unp'] = $g_{\mu\nu}$
gLow['sym'] = defPertSum(gLow['unp'],gPertList) # Symbolic decomposition
gLow['ord'] = defPertList($h_{\mu\nu}$,pertLabel,maxPertOrder) # Initialize (momentarily) ord as a pertList of h symbols
gLow['ord'][0] = $\eta_{\mu\nu}$
for i in range(maxPertOrder+1): 
  l = gPertList[i]
  r = gLow['ord'][i]
  gLow['ord'][i] = $@(l) = @(r)$
\end{cadabra}
We highlight the use of out function \texttt{getPertList}, which
assigns to \verb+gLow['ord']+ a alist with all the perturbative
orders characterized by a weight proportional to the array position
and labelled by \texttt{pert}. All the properties of $h_{\mu\nu}$ are
inherited inside the $\p{i}{h}$ by the same function. The function
\texttt{defPertSum} sums all the perturbative orders into
\verb+gLow['sym']+.

In order to test the effectiveness of the code, we try to print the
contents of the  dictionary 
\begin{cadabra}[numbers=none]
gLow['unp'];
gLow['sym'];
gLow['ord'][1];
\end{cadabra}
getting respectively
\begin{dgroup*}[noalign]
  \begin{dmath*}
    g_{\mu \nu}
  \end{dmath*}
  \begin{dmath*}
    g_{\mu \nu} = \p{0}{g_{\mu \nu}}+\p{1}{g_{\mu \nu}}+\p{2}{g_{\mu
        \nu}}
  \end{dmath*}
  \begin{dmath*}
    \p{1}{h_{\mu \nu}}
  \end{dmath*}
\end{dgroup*}
Note that our algorithm for \verb+gLow+ inherits properties of
$g_{\mu\nu}$ on $\p{i}{h_{\mu\nu}}$ independentely to the position of
their indices position. Consequentely, part of the command used can be
already employed to calculate the contravariant metric perturbations. 

\subsection{Contravariant Metric Perturbations \textnormal{\texttt{gUpp[]}}}
\label{sec:Metup}

Similarly to above we can define the \verb+gUpp+ object with entries
\verb+unp+, \verb+sym+ and \verb+ord+. Unlike before, the definition
of the \verb+gUpp[]+ dictionary for covariant metric perturbations is
a slightly more complicated, as they must defined in order to satisfy
the following identity at each perturbative order:
\begin{equation}
  \label{eq:metrid}
  g_{\mu\alpha}g^{\alpha\nu}=\delta^\mu_\nu\,,
\end{equation}
with $\delta^\mu_\nu$ the Kr\"{o}necker delta function that has a
background \texttt{WeightInherit} propriety defined in Section
\ref{sec:env}.  

We begin with the same approach of the previous case and load
temporarily \texttt{gUpp[]} as:
\StartLineAt{23}
\begin{cadabra}
gUpp = {} 
gUpp['unp'] = $g^{\mu\nu}$
gUpp['sym'] = defPertSum(gUpp['unp'],defPertList(gUpp['unp'],pertLabel,maxPertOrder))
gUpp['ord'] = [None]*(maxPertOrder+1) # Initialize (momentarily) ord as empty list of fixed positions
gUpp['ord'][0] = $g0^{\mu\nu}=\eta^{\mu\nu}$
\end{cadabra}
Notice that the \verb+defPertList()+ function is called inside
\verb+defPertSum+. We initialized the \verb+ord+ as a list of
\verb+None+ values, except the background to eta. Note the use of
\verb+g0+, defined and available thanks to the use of the defPertList
command used for $g_{\mu\nu}$ in previous code. Note that in this case
we directly deal with equalities for \verb+ord+.

So far, we have that \verb+gUpp[]+ has the same content of
\texttt{gLow[]} with raised indices. Some of these values have only
been provisionally uploaded and they will be soon replaced. Now, let
us introduce the identity \eqref{eq:metrid} written in a slightly
modified form
\ContinueLineNumber
\begin{cadabra}
ident := 0 = \delta_{\rho}^{\nu}-g_{\rho\alpha}g^{\alpha\nu};
\end{cadabra}
to calculate the entries. 

This identity can be written in terms of the perturbative expansions
held in \verb|gLow['sym']| and \verb|gUpp['sym']|. Let us substitute
inside it all the obtained decompositions \verb+gLow['dec']+ and
\verb+gUpp['dec']+ with our \texttt{subsPertSums()} function of
Subsection \ref{sec:func1},
\begin{cadabra}
subsPertSums(ident,pertLabel,maxPertOrder,gLow['sym'],gUpp['sym']);
\end{cadabra}
\begin{gather}
  \label{eq:mas}
  0 = \delta_{\rho}{}^{\nu}-\p{0}{g_{\rho \alpha}} \p{0}{g^{\alpha \nu}}-\p{0}{g_{\rho \alpha}} \p{1}{g^{\alpha \nu}}-\p{0}{g_{\rho \alpha}} \p{2}{g^{\alpha \nu}}-\p{1}{g_{\rho \alpha}} \p{0}{g^{\alpha \nu}}-\p{1}{g_{\rho \alpha}} \p{1}{g^{\alpha \nu}}-\p{2}{g_{\rho \alpha}} \p{0}{g^{\alpha \nu}}
\end{gather}
Note that unnecessary high order terms like $\p{2}{g_{\rho \alpha}}
\p{2}{g^{\alpha \nu}}$ are removed, since \verb+maxPertOrder=2+.  

As $\p{0}{g_{\mu\nu}} = \eta_{\mu\nu}$ it is clear that
$\p{0}{g_{\rho\alpha}}\p{0}{g^{\alpha\nu}} = \delta_\rho^{~\nu}$ and
so we can find $\p{i}{g}^{\mu\nu}$ by looking at the $i$-th term
in the expansion, which is what the function
\verb|getEquationpertOrder| from the \verb|libraries.perturbations|
nortebook does. Examining only terms of order $i$, the quantity
$\p{i}{g^{\alpha\nu}}$ can only be accompanied by the factor
$\p{0}{g_{\rho\alpha}}$ which is the flat metric. By contracting with
$\eta^{\mu\rho}$ we can isolate this quantity and then write it in
terms of the lower order perturbations by substituting in for the
values already calculated in \verb|gLow['ord']| and
\verb|gUpp['ord']|. Following such procedure, using Eq.~\eqref{eq:mas}
we can assign the correct values to \verb+gUpp['ord']+ as follows:
\begin{cadabra}
for i in range(1,maxPertOrder+1):
  # Select the ith order perturbation
  order_i = getEquationPertOrder(ident,pertLabel,i)
  # Multiply through by g0^{\mu\rho} and move gi^{\alpha\nu} to the other side
  manip.multiply_through(order_i, $g0^{\mu\rho}$)
  distribute(order_i)
  substitute(order_i, $g0^{\mu\rho}g0_{\rho\alpha}->\delta^{\mu}_{\alpha}$)
  manip.add_through(order_i, Ex(r'\delta^{\mu}_{\alpha}'+r'g'+str(i)+r'^{\alpha\nu}'))
  # Substitute definitions from gLow['ord'] and gUpp['ord']
  for j in range(0,i+1):
    substitute(order_i,gLow['ord'][j]) 
  for j in range(0,i):  
    substitute(order_i,gUpp['ord'][j])
  # Simplify result and assign to gUpp['ord']
  distribute(order_i)
  eliminate_kronecker(order_i,repeat=True)
  eliminate_metric(order_i,repeat=True)
  canonicalise(order_i)
  gUpp['ord'][i] = order_i
\end{cadabra}

In order to test the effectiveness of the code, we print the content of the dictionary
\begin{cadabra}[numbers=none]
gUpp['unp'];
gUpp['sym'];
gUpp['ord'][1];
\end{cadabra}
\begin{dgroup*}[noalign]
  \begin{dmath*}
    g^{\mu \nu}
  \end{dmath*}
  \begin{dmath*}
    g^{\mu \nu} = \p{0}{g}^{\mu \nu} + \p{1}{g}^{\mu \nu} + \p{2}{g}^{\mu \nu}
  \end{dmath*}
  \begin{dmath*}
    \p{1}{g}^{\mu \nu} = - \p{1}{h}^{\mu \nu}
  \end{dmath*}
\end{dgroup*}

\subsection{Generalized Perturbative Function}
\label{sec:genpert}

As we have discussed in the previous sections, working with metric
tensor perturbations has in a naturally way introduced a
dictionary-model approach in order to collect all the obtained
perturbative informations. Following the same philosophy, we
modularized our coding in order to highlight the main logical
building-blocks operations to decompose a generic tensor quantity. In
fact, up to some little modifications, all operations previously
discussed can be chained automatically to define a generic
decomposition function that we call \texttt{perturb}.

\boxedFunction{perturb(ex:Ex, pertDicts:Dict[Ex], pertLabel:str, maxPertOrder:str) -> Dict[Ex]}
{Take an \texttt{Ex} object and produces with respect to it a Python
  dictionary containing the valorized perturbative structure defined
  in Table \ref{tab:sometab}.}

The \verb|perturb| function below takes an \verb|Ex| object and
produces dictionaries similar to \verb|gUpp| and \verb|gLow|. Note
that the \verb|ord| entry is only calculated when the input \verb|ex|
is an equation.
\StartLineAt{49}
\begin{cadabra}
def perturb(ex, pertDicts, pertLabel, maxPertOrder):
  # Initialize the pertDict object which will be returned
  pertDict = {}
  pertDict['ord'] = [None]*(maxPertOrder+1)
  if ex.top().name != r"\equals":
    # Not an equation, 'unp' is just ex and 'sym' the result of defPertSum
    pertDict['unp'] = ex
    pertDict['sym'] = defPertSum(ex,defPertList(ex,pertLabel,maxPertOrder))
  else:
    # 'unp' and 'sym' are as above
    pertDict['unp'] = ex[0]
    pertDict['sym'] = defPertSum(ex[0],defPertList(ex[0],pertLabel,maxPertOrder))
    # If pertDicts are provided, use them to 
    if pertDicts is not None:
      # Decompose pertDicts into two components for 'sym' and 'ord'
      symPertList=[]
      ordPertList=[]
      for dic in pertDicts: 
        symPertList.append(dic['sym'])
        ordPertList.append(dic['ord'])
      # Substitute symbolic decompositions
      subbed_ex = subsPertSums($@(ex)$, pertLabel, maxPertOrder, *symPertList, pertDict['sym'])
      # Iterate over all orders substituting order decompositions
      for i in range(0,maxPertOrder+1):
        cur_order = getEquationPertOrder(subbed_ex, pertLabel, i)
        for dec in ordPertList: # All objects to substitute
          if dec[0] is not None: # Leave symbolic the expressions
            for j in range(i+1): # All orders up to i
              substitute(cur_order, dec[j], repeat=True)
        # Simplifications
        distribute(cur_order)
        unwrap(cur_order,repeat=True)
        product_rule(cur_order)
        eliminate_metric(cur_order,repeat=True)
        eliminate_kronecker(cur_order,repeat=True)
        # Assign result of calculation to 'ord' 
        pertDict['ord'][i] = cur_order
  return pertDict
\end{cadabra}

In addition to the \verb+ex+ object to be perturbed and the standard
setting arguments \verb+pertLabel+ and \verb+maxPertOrder+, the
\verb+perturb()+ function requires the \verb+pertDicts+ variable,
namely the list perturbative units needed to perturb \verb+ex+
organized in terms of dictionaries whose structure is described in
Table \ref{tab:sometab}. Such perturbative units can be obtained by
applying the \verb+perturb()+ function itself to the lower levels of
the perturbative chain conducting to \verb+ex+. As an example, as
shown in Subsection \ref{sec:pertsoth}, the perturbative expansion of
the Einstein tensor $G_{\mu\nu}(g_{\mu\nu}, R_{\mu\nu}, R)$ can be
obtained from the perturbative chain
$g_{\mu\nu}\rightarrow g^{\mu\nu}\rightarrow
\Gamma^{\sigma}_{\mu\nu}\rightarrow R_{\mu\nu\rho\sigma}\rightarrow
R_{\mu\nu}\rightarrow R\rightarrow G_{\mu\nu}$. Then, the passed
\verb+pertDicts+ are decomposed into the two lists \verb+symPertList+
and \verb+ordPertList+, collecting all \verb+'sym'+ and \verb+'ord'+
contributes separately. Such lists are then employed as substitutions
into \verb+cur_order+, the symbolic decomposition of \verb+ex+ for
each perturbative order. The last part of the function is dedicated to
simplifications using standard \verb+Cadabra+ commands. In particular, for
optimization reasons, the use of the burdensome \verb+canonicalise+
algorithm was deliberately avoided.

\subsubsection{Tensor Perturbations: illustrative calculations}
\label{sec:pertsoth}

We are now ready to concretely test the tools developed in the
previous sections, in particular the use of the \texttt{perturb}
function. The following code creates in few line commands the tensor
decomposition of the main metric objects of General Relativity passed
from the \emph{header.cnb} file, giving rise to a powerful approach to
deal with higher-order perturbations within the \verb+Cadabra+ and
Python framework:
\ContinueLineNumber
\begin{cadabra}
connection = perturb(ch(), [gLow, gUpp], pertLabel, maxPertOrder)
riemann = perturb(rm(), [connection], pertLabel, maxPertOrder)
ricciTensor = perturb(rc(), [riemann], pertLabel, maxPertOrder)
ricciScalar = perturb(rs(), [gUpp,ricciTensor], pertLabel, maxPertOrder)
einstein = perturb(ei(), [gLow,ricciTensor,ricciScalar], pertLabel, maxPertOrder)
\end{cadabra}
To verify if the code works correctly, we try to print the
\texttt{ricciTensor} content: 
\begin{cadabra}[numbers=none]
ricciTensor['unp'];
ricciTensor['sym'];
ricciTensor['ord'][1];
\end{cadabra}
which finally gives, as expected:
\begin{dgroup*}[noalign]
  \begin{dmath*}
    R_{\sigma \nu}
  \end{dmath*}
  \begin{dmath*}
    R_{\sigma \nu} = \p{0}{R}_{\sigma \nu}+\p{1}{R}_{\sigma
      \nu}+\p{2}{R}_{\sigma \nu}
  \end{dmath*}
  \begin{dmath*}
    \p{1}{R}_{\nu \sigma} = \tfrac{1}{2}\partial_{\sigma
      \tau}{\p{1}{h}_{\nu}\,^{\tau}} -
    \tfrac{1}{2}\partial^{\tau}\,_{\tau}{\p{1}{h}_{\nu \sigma}} -
    \tfrac{1}{2}\partial_{\nu
      \sigma}{\p{1}{h}_{\tau}\,^{\tau}}+\tfrac{1}{2}\partial_{\nu}\,^{\tau}{\p{1}{h}_{\sigma
        \tau}}
  \end{dmath*}
\end{dgroup*}

A further tensor quantity that will prove extremely important in the
following sections is the so-called harmonic gauge condition. The
harmonic gauge condition is defined as
\begin{equation}
  \label{eq:harm}
  \Gamma^{\sigma} = g^{\mu \nu} \Gamma^{\sigma}\,_{\mu \nu}\,.
\end{equation}
Using \verb+Cadabra+ language we can write:
\StartLineAt{92}
\begin{cadabra}
hg{#}::LaTeXForm("\Gamma").
hg := hg^{\sigma} = g^{\mu\nu} ch^{\sigma}_{\mu\nu};
\end{cadabra}
Following the same approach of the previous decompositions, we compute
\begin{cadabra}
harmonicGauge = perturb(hg,[gUpp,connection],pertLabel,maxPertOrder)  
\end{cadabra}
yielding 
\begin{cadabra}[numbers=none]
harmonicGauge['unp'];
harmonicGauge['sym'];
harmonicGauge['ord'][2];
\end{cadabra}
\begin{dgroup*}[noalign]
  \begin{dmath*}
    \Gamma^{\sigma}
  \end{dmath*}
  \begin{dmath*}
    \Gamma^{\sigma} =
    \p{0}{\Gamma}^{\sigma}+\p{1}{\Gamma}^{\sigma}+\p{2}{\Gamma}^{\sigma}
  \end{dmath*}
  \begin{dmath*}
    \p{2}{\Gamma}^{\sigma} =
    \tfrac{1}{2}\partial_{\nu}{\p{2}{h}^{\nu
        \sigma}}+\tfrac{1}{2}\partial^{\nu}{\p{2}{h}_{\nu}\,^{\sigma}}
    - \tfrac{1}{2}\partial^{\sigma}{\p{2}{h}^{\nu}\,_{\nu}} -
    \tfrac{1}{2}\p{1}{h}^{\sigma \nu}
    \partial_{\rho}{\p{1}{h}^{\rho}\,_{\nu}} -
    \tfrac{1}{2}\p{1}{h}^{\sigma \nu} \partial^{\rho}{\p{1}{h}_{\nu
        \rho}}+\tfrac{1}{2}\p{1}{h}^{\sigma \nu}
    \partial_{\nu}{\p{1}{h}^{\rho}\,_{\rho}} -
    \tfrac{1}{2}\p{1}{h}^{\mu \nu}
    \partial_{\nu}{\p{1}{h}_{\mu}\,^{\sigma}} -
    \tfrac{1}{2}\p{1}{h}^{\mu \nu}
    \partial_{\mu}{\p{1}{h}_{\nu}\,^{\sigma}}+\tfrac{1}{2}\p{1}{h}^{\mu
      \nu} \partial^{\sigma}{\p{1}{h}_{\mu \nu}}
  \end{dmath*}
\end{dgroup*}

We can also deal with matter sources like \emph{energy-momentum}
tensor and its trace:
\StartLineAt{93}
\begin{cadabra}
mt := T_{\mu\nu}:
tr := T = g^{\mu\nu}T_{\mu\nu}:
#Properties
{T_{\mu\nu},T_{\mu\nu}}::Symmetric.
T{#}::Depends(\partial{#}).
\end{cadabra}
Using the \verb+perturb()+ function we calculate
\ContinueLineNumber
\begin{cadabra}
matter = perturb(mt,None,pertLabel,maxPertOrder)
\end{cadabra}
to get
\begin{cadabra}[numbers=none]
matter['unp'];
matter['sym'];
matter['ord'][2];
\end{cadabra}
\begin{dgroup*}[noalign]
  \begin{dmath*}
    T_{\mu \nu}
  \end{dmath*}
  \begin{dmath*}
    T_{\mu \nu} = \p{0}{T}_{\mu \nu}+\p{1}{T}_{\mu \nu}+\p{2}{T}_{\mu \nu}
  \end{dmath*}
  \begin{dmath*}
    \mbox{\texttt{None}}
  \end{dmath*}
\end{dgroup*}
Note that if we never pass the list of perturbative dictionaries, a
\verb+None+ is returned for \verb+ord+. 

An additional useful example is the following, where a mixture of
\verb+sym+ and \verb+ord+ is passed to perturb the trace of the
energy-momentum tensor:
\StartLineAt{99}
\begin{cadabra}
matterTr = perturb(tr,[gUpp,matter],pertLabel,maxPertOrder)
\end{cadabra}
\begin{cadabra}[numbers=none]
matterTr['unp'];
matterTr['sym'];
matterTr['ord'][2];
\end{cadabra}
\begin{dgroup*}[noalign]
  \begin{dmath*}
    T
  \end{dmath*}
  \begin{dmath*}
    T = \p{0}{T}+\p{1}{T}+\p{2}{T}
  \end{dmath*}
  \begin{dmath*}
    \p{2}{T} = \p{2}{T}^{\nu}\,_{\nu} - \p{1}{h}^{\mu \nu}
    \p{1}{T}_{\mu \nu} + \p{1}{h}^{\alpha}\,_{\mu} \p{1}{h}^{\mu \nu}
    \p{0}{T}_{\alpha \nu} - \p{2}{h}^{\mu \nu} \p{0}{T}_{\mu \nu}
  \end{dmath*}
\end{dgroup*}
It is clear that the tools developed in this Section can be completely
customized according to need and used for obtaining higher
perturbative orders of the objects already analyzed, just changing the
value of \texttt{maxPertOrder}, or for the decomposition of any other
tensor.

\section{First-order gravitational-waves equation in presence of matter}
\label{sec:fistgw}

At this point, we have the fundamental elements for the perturbative
decomposition of relevant physical equations. In this Section
\ref{sec:fistgw} we linearize the Einstein equations in order to
explicitly derive, within the harmonic gauge \eqref{eq:harm}, the
wave-equation which describes the propagation of first-order
gravitational waves in presence of matter.

For this example we require the results from the previous tutorial
\verb+sec4_TensorPerturbationsGR+, and the \verb+header+ module is
imported by inheritance.
\StartLineAt{1}
\begin{cadabra}
from sec4_TensorPerturbationsGR import *
\end{cadabra}
With the previous tutorials we have the fundamental elements for the
perturbative decomposition of the Einstein's equations. More details
on how obtaining these equations can be found in the tutorial
\emph{Einstein equations from a variational principle}.\footnote{The
  tutorial is available in the web page \url{https://cadabra.science/notebooks/einstein_equations.html}}

\ContinueLineNumber
\begin{cadabra}
einEq := ei_{\mu\nu}=\kappa T_{\mu\nu};
substitute(_, ei());
\end{cadabra}
\begin{dgroup*}[noalign]
  \begin{dmath*}
    G_{\mu \nu} = \kappa T_{\mu \nu}
  \end{dmath*}
  \begin{dmath*}
    R_{\mu \nu} - \tfrac{1}{2}g_{\mu \nu} R = \kappa T_{\mu \nu}
  \end{dmath*}
\end{dgroup*}
We remember that $\kappa$ is a constant associated to the
gravitational coupling. Einstein's equations can be reformulated by
taking the trace of the equations:
\begin{cadabra}
# Multiply through by the metric
trEinEq := @(einEq);
manip.multiply_through(_, $g^{\mu\nu}$)
distribute(_);
# Replace contracted indices with scalar quantities
substitute(_, $g^{\mu\nu}T_{\mu\nu}->T$)
substitute(_, manip.eq_to_subrule(rs()))
eliminate_metric(_)
eliminate_kronecker(_)
# Clean up
manip.multiply_through(_,$-1$); 
\end{cadabra}
\begin{dgroup*}[noalign]
  \begin{dmath*}
    R_{\mu \nu} - \tfrac{1}{2}g_{\mu \nu} R = \kappa T_{\mu \nu}
  \end{dmath*}
  \begin{dmath*}
    g^{\mu \nu} R_{\mu \nu} - \tfrac{1}{2}g^{\mu \nu} g_{\mu \nu} R = g^{\mu \nu} \kappa T_{\mu \nu}
  \end{dmath*}
  \begin{dmath*}
    R = -T \kappa
  \end{dmath*}
\end{dgroup*}

We can now use the expressions \verb|einEq| and \verb|trEinEq| to
write out the source term $S_{\mu\nu}$ which we define here using the
\verb|perturb| function from the pervious section
\begin{cadabra}
S_{\mu\nu}::Symmetric.
sm := S_{\mu\nu} = T_{\mu\nu} - 1/2 g_{\mu\nu} T;
source = perturb(_,[gLow,matter,matterTr],pertLabel,maxPertOrder)
\end{cadabra}
\begin{dmath*}
S_{\mu \nu} = T_{\mu \nu} - \frac{1}{2}g_{\mu \nu} T
\end{dmath*}

It should be noted that for the source tensor $S_{\mu\nu}$ we decide
to use only the symbolic decomposition, labelled by the key name
\verb+'sum'+ in the dictionary \texttt{source}. Printing the result
we get
\begin{cadabra}
source['sym'];
source['ord'][0];
\end{cadabra}
\begin{dgroup*}
  \begin{dmath*}
    S_{\mu \nu} = \p{0}{S}_{\mu \nu}+\p{1}{S}_{\mu \nu}+\p{2}{S}_{\mu \nu}
  \end{dmath*}
  \begin{dmath*}
    \p{0}{S}_{\mu \nu} = \p{0}{T}_{\mu \nu} - \tfrac{1}{2}\eta_{\mu \nu} \p{0}{T}^{\rho}\,_{\rho}
  \end{dmath*}
\end{dgroup*}
In terms of this quantity the Einstein equations become
\begin{cadabra}
einEq2 := @(einEq):
manip.isolate(_, $rc_{\mu\nu}$);
# Insert expressions for source term and scalar Einstein equation
substitute(_, manip.isolate(sm, $T_{\mu\nu}$))
substitute(_, trEinEq);
# Clean up
distribute(_)
sort_product(_);
\end{cadabra}
\begin{dgroup*}[noalign]
  \begin{dmath*}
    R_{\mu \nu} = \kappa T_{\mu \nu}+\tfrac{1}{2}g_{\mu \nu} R
  \end{dmath*}
  \begin{dmath*}
    R_{\mu \nu} = \kappa \left(\tfrac{1}{2}g_{\mu \nu} T+S_{\mu \nu}\right) - \tfrac{1}{2}g_{\mu \nu} T \kappa
  \end{dmath*}
  \begin{dmath}
    \label{eq:previous}
    R_{\mu \nu} = S_{\mu \nu} \kappa
  \end{dmath}
\end{dgroup*}
Equation \eqref{eq:previous} is the same one obtained in our previous
work
\cite{castillo-felisola20_cadab_python_algor_gener_relat_cosmol_i} for
$\Lambda=0$ and the definition of $S_{\mu\nu}$.

The first-order perturbed component of the above expression is given by
\begin{cadabra}
einEq2 = subsPertSums(einEq2,pertLabel,maxPertOrder,source['sym'],ricciTensor['sym']);
einEq2 = getEquationPertOrder(einEq2,pertLabel,1);
substitute(_,ricciTensor['ord'][1]);
\end{cadabra}
\begin{dgroup*}[noalign]
  \begin{dmath*}
    R_{\mu \nu}+\p{1}{R}_{\mu \nu}+\p{2}{R}_{\mu \nu} = \p{0}{S}_{\mu \nu} \kappa+\p{1}{S}_{\mu \nu} \kappa+\p{2}{S}_{\mu \nu} \kappa
  \end{dmath*}
  \begin{dmath*}
    \p{1}{R}_{\mu \nu} = \p{1}{S}_{\mu \nu} \kappa
  \end{dmath*}
  \begin{dmath*}
    \tfrac{1}{2}\partial_{\nu \tau}{\p{1}{h}_{\mu}\,^{\tau}} - \tfrac{1}{2}\partial^{\tau}\,_{\tau}{\p{1}{h}_{\mu \nu}} - \tfrac{1}{2}\partial_{\mu \nu}{\p{1}{h}_{\tau}\,^{\tau}}+\tfrac{1}{2}\partial_{\mu}\,^{\tau}{\p{1}{h}_{\nu \tau}} = \p{1}{S}_{\mu \nu} \kappa
  \end{dmath*}
\end{dgroup*}
This equation does not permit a unique solution since given any
solution it will always be possible to identify another solution by
performing a coordinate transformation. This property is known as as
\emph{gauge invariance}. This redundancy can be removed by fixing a
specific coordinate system: for our purposes it is a good choice to
work in the \emph{harmonic gauge}, defined by the condition:
\begin{cadabra}
hgCond := 0 = hg^{\sigma};
\end{cadabra}
\begin{dmath*}
0 = \Gamma^{\sigma}
\end{dmath*}
The first-order perturbed component of the harmonic gauge condition is
\begin{cadabra}
hgCond1 := @(hgCond):
hgCond1 = subsPertSums(_,pertLabel,maxPertOrder,harmonicGauge['sym']);
hgCond1 = getEquationPertOrder(_,pertLabel,1);
substitute(_,harmonicGauge['ord'][1]);
# Combine terms which only differ in height on contracted dummy indices
hgCond1 = substitute(_,$\partial_{\nu}{h1^{\nu\sigma}}->\partial^{\nu}{h1_{\nu}^{\sigma}}$);
\end{cadabra}
\begin{dgroup*}[noalign]
  \begin{dmath*}
    0 = \p{0}{\Gamma}^{\sigma}+\p{1}{\Gamma}^{\sigma}+\p{2}{\Gamma}^{\sigma}
  \end{dmath*}
  \begin{dmath*}
    0 = \p{1}{\Gamma}^{\sigma}
  \end{dmath*}
  \begin{dmath*}
    0 = \tfrac{1}{2}\partial_{\nu}{\p{1}{h}^{\nu \sigma}}+\tfrac{1}{2}\partial^{\nu}{\p{1}{h}_{\nu}\,^{\sigma}} - \tfrac{1}{2}\partial^{\sigma}{\p{1}{h}^{\nu}\,_{\nu}}
  \end{dmath*}
  \begin{dmath*}
    0 = \partial^{\nu}{\p{1}{h}_{\nu}\,^{\sigma}} - \tfrac{1}{2}\partial^{\sigma}{\p{1}{h}^{\nu}\,_{\nu}}
  \end{dmath*}
\end{dgroup*}
This allows us to obtain the following constraint
\begin{cadabra}
manip.multiply_through(hgCond1, $2*\eta_{\alpha\sigma}$)
manip.apply_through(_, $\partial_{\beta}{#}$);
# Calculate derivatives
distribute(_)
product_rule(_)
unwrap(_);
# Eliminate metric and canonicalise
eliminate_metric(_)
canonicalise(_)
eliminate_kronecker(_)
rename_dummies(_);
\end{cadabra}
\begin{dgroup*}[noalign]
  \begin{dmath*}
    0 = 2\partial_{\beta}\left[\eta_{\alpha \sigma} \left(\partial^{\nu}{\p{1}{h}_{\nu}\,^{\sigma}} - \tfrac{1}{2}\partial^{\sigma}{\p{1}{h}^{\nu}\,_{\nu}}\right)\right]
  \end{dmath*}
  \begin{dmath*}
    0 = 2\eta_{\alpha \sigma} \partial_{\beta}\,^{\nu}{\p{1}{h}_{\nu}\,^{\sigma}}-\eta_{\alpha \sigma} \partial_{\beta}\,^{\sigma}{\p{1}{h}^{\nu}\,_{\nu}}
  \end{dmath*}
  \begin{dmath*}
    0 = 2\partial_{\beta}\,^{\mu}{\p{1}{h}_{\alpha \mu}}-\partial_{\alpha \beta}{\p{1}{h}^{\mu}\,_{\mu}}
  \end{dmath*}
\end{dgroup*}

In the following code, the substitution $h^\mu_\mu \to h$ is made to
allow the operator \texttt{manip.to_lhs} the automatically recognise
that the addend should be moved to the left hand side of the equation. 
\begin{cadabra}
substitute(_,$h1^{\mu}_{\mu}->h1$);
manip.to_lhs(_,$\partial_{\alpha \beta}(h1)$, $exact=True$);
substitute(_,$h1->h1^{\mu}_{\mu}$);
\end{cadabra}
\begin{dgroup*}[noalign]
  \begin{dmath*}
    0 = 2\partial_{\beta}\,^{\mu}{\p{1}{h}_{\alpha \mu}}-\partial_{\alpha \beta}{\p{1}{h}}
  \end{dmath*}
  \begin{dmath*}
    \partial_{\alpha \beta}{\p{1}{h}} = 2\partial_{\beta}\,^{\mu}{\p{1}{h}_{\alpha \mu}}
  \end{dmath*}
  \begin{dmath*}
    \partial_{\alpha \beta}{\p{1}{h}^{\mu}\,_{\mu}} = 2\partial_{\beta}\,^{\mu}{\p{1}{h}_{\alpha \mu}}
  \end{dmath*}
\end{dgroup*}
As the partial derivatives commute, we can also deduce that
\begin{cadabra}
\partial_{\beta}^{\mu}(h1_{\alpha \mu})::Symmetric.
\end{cadabra}
Introducing the gauge conditions just obtained in the perturbed
Einstein equations \texttt{gw_pert}, some suitable algebraic
manipulations provide the well-known wave equation in the presence of
sources
\begin{cadabra}
einEqGauge = manip.multiply_through(einEq2, $-2$)
substitute(einEqGauge, $h1_{\mu}^{\mu}->h1^{\mu}_{\mu}$);
substitute(_, hgCond1);
substitute(_, $\partial_{\nu\rho}(h1_{\mu}^{\rho}) -> \partial_{\mu\nu}(h1^{\rho}_{\rho})$)
canonicalise(_);
\end{cadabra}
\begin{dgroup*}[noalign]
  \begin{dmath*}
    -\partial_{\nu \tau}{\p{1}{h}_{\mu}\,^{\tau}}+\partial^{\tau}\,_{\tau}{\p{1}{h}_{\mu \nu}}+\partial_{\mu \nu}{\p{1}{h}^{\tau}\,_{\tau}}-\partial_{\mu}\,^{\tau}{\p{1}{h}_{\nu \tau}} = -2\p{1}{S}_{\mu \nu} \kappa
  \end{dmath*}
  \begin{dmath*}
    -\partial_{\nu \tau}{\p{1}{h}_{\mu}\,^{\tau}}+\partial^{\tau}\,_{\tau}{\p{1}{h}_{\mu \nu}}+2\partial_{\nu}\,^{\tau}{\p{1}{h}_{\mu \tau}}-\partial_{\mu}\,^{\tau}{\p{1}{h}_{\nu \tau}} = -2\p{1}{S}_{\mu \nu} \kappa
  \end{dmath*}
  \begin{dmath}
    \boxed{\partial^{\tau}\,_{\tau}{\p{1}{h}_{\mu \nu}} = -2\p{1}{S}_{\mu \nu} \kappa}\qquad\mbox{(Wave Equation)}
  \end{dmath}
\end{dgroup*}
In absence of matter fields we get
\begin{cadabra}
waveEq = substitute(_, $S1_{\mu\nu}->0$);
\end{cadabra}
\begin{equation}\label{eq:nmn}
\partial^{\tau}\,_{\tau}{\p{1}{h}_{\mu \nu}} = 0
\end{equation}
The above equation shows that every component of the metric
perturbation satisfy the wave equation. 

\section{First-order gravitational wave-relations}
\label{sec:gwrel}

Working in the so-called harmonic gauge (see Eq. \eqref{eq:harm}), we
consider plane-wave solutions of the wave equation~\eqref{eq:nmn}
which describes the propagation of gravitational waves in absence of
matter fields. We show that such solutions must satisfy the well-known
wave-relations. This section is useful for showing the versatility of
\texttt{cadabra} in managing tensor expressions that belong to the
argument of exponential functions.

As with the notebook from the previous section we begin by importing
the results of section \ref{sec:tenspert}, on top of which we import
two functions from the \verb|cadabra| standard library:
\verb|replace_index| which renames all indices in a subexpression, and
\verb|get_basis_component| which extracts the coefficient of a term.
\StartLineAt{1}
\begin{cadabra}
from sec4_TensorPerturbationsGR import *
from cdb.core.manip import get_basis_component
from cdb.utils.indices import replace_index
\end{cadabra}

\subsection{Definitions}

We introduce the spacetime coordinate $x^\mu$, the polarization tensor
$\mathbf{e}_{\mu\nu}$ and its complex conjugate
$\bar{\mathbf{e}}_{\mu\nu}$
\ContinueLineNumber
\begin{cadabra}
x::Coordinate.
x{#}::Depends(\partial{#}).
\pol1{#}::LaTeXForm("\mathbf{e}").
\pol2{#}::LaTeXForm("\bar{\mathbf{e}}").
\end{cadabra}
Now we can consider the plane-wave solution, parameterised by the wave-number $k_\lambda$,
\begin{cadabra}
sol:=h1_{\mu\nu}=\pol1_{\mu\nu}\exp(i k_\lambda x^{\lambda}) + \pol2_{\mu\nu}\exp(-i k_\lambda x^{\lambda});
\end{cadabra}
\begin{dmath*}
  \p{1}{h}_{\mu \nu} = \mathbf{e}_{\mu \nu} \exp\left(i k_{\lambda}
    x^{\lambda}\right)+\bar{\mathbf{e}}_{\mu \nu} \exp\left(-i
    k_{\lambda} x^{\lambda}\right)
\end{dmath*}
where $i$ is the imaginary unit. Here $h_{\mu\nu}$ is symmetric and
inherits its dependence on the derivative $\partial_\mu$ through
$\exp(\pm i k_\lambda x^{\lambda})$
\begin{cadabra}
\exp(i k_\lambda x{#})::Depends(\partial{#}).
\exp(-i k_\lambda x{#})::Depends(\partial{#}).
\end{cadabra}
For the purposes of our analysis, we restrict to the vacuum
wave-equation 
\begin{cadabra}
wave_eq := \partial_{\sigma}{\partial^{\sigma}{h1_{\mu \nu}}} = 0;
\end{cadabra}
\begin{dmath*}
  \partial_{\sigma}\,^{\sigma}{\p{1}{h}_{\mu \nu}} = 0
\end{dmath*}
and by the harmonic gauge condition
\begin{cadabra}
gauge_eq := \partial_{\mu}{h1^{\mu}_{\nu}}} - 1/2*\partial_{\nu}{h1^{\mu}_{\mu}} = 0;
\end{cadabra}
\begin{dmath*}
  \partial_{\mu}{\p{1}{h}^{\mu}\,_{\nu}} -
  \tfrac{1}{2}\partial_{\nu}{\p{1}{h}^{\mu}\,_{\mu}} = 0
\end{dmath*}
More details on obtaining the wave-equation and the harmonic gauge
first-order decomposition can be found in the previous tutorials. In
order to handle the derivatives of the exponential function
$\exp(\pm i k_\lambda x^{\lambda})$, the following replacement rules
are required:
\begin{cadabra}
rule_1:=\partial_{\sigma}^{\sigma}(\exp(A??))->\partial_{\sigma}{\exp(A??)*\partial^{\sigma}{A??}};
rule_2:=\partial^{\mu}{\exp(A??)}->\exp(A??)*\partial^{\mu}{A??};
rule_3:=\partial_{\mu}{\exp(A??)}->\exp(A??)*\partial_{\mu}{A??};
\end{cadabra}
\begin{dgroup*}[noalign]
  \begin{dmath*}
    \partial_{\sigma}\,^{\sigma}\left(\exp\left(A??\right)\right)
    \rightarrow \partial_{\sigma}\left(\exp\left(A??\right)
      \partial^{\sigma}\left(A??\right)\right)
  \end{dmath*}
  \begin{dmath*}
    \partial^{\mu}\left(\exp\left(A??\right)\right) \rightarrow
    \exp\left(A??\right) \partial^{\mu}\left(A??\right)
  \end{dmath*}
  \begin{dmath*}
    \partial_{\mu}\left(\exp\left(A??\right)\right) \rightarrow
    \exp\left(A??\right) \partial_{\mu}\left(A??\right)
  \end{dmath*}
\end{dgroup*}

\subsection{First wave-relation}

We begin by imposing that the plane-wave $h_{\mu\nu}$ is a solution of
the vacuum wave equation 
\begin{cadabra}
wave_eq1 := @(wave_eq);
substitute(_, sol);
\end{cadabra}
\begin{dgroup*}[noalign]
  \begin{dmath*}
    \partial_{\sigma}\,^{\sigma}{\p{1}{h}_{\mu \nu}} = 0
  \end{dmath*}
  \begin{dmath*}
    \partial_{\sigma}\,^{\sigma}\left(\mathbf{e}_{\mu \nu} \exp\left(i
        k_{\lambda} x^{\lambda}\right)+\bar{\mathbf{e}}_{\mu \nu}
      \exp\left(-i k_{\lambda} x^{\lambda}\right)\right) = 0
  \end{dmath*}
\end{dgroup*}
To simplify the expression, we need to use the \verb+Cadabra+ specific
\verb|converge| construction, which is essentially a while loop which
runs until the predicate expression no longer changes
\begin{cadabra}
converge(wave_eq1):
  distribute(_)
  product_rule(_)
  unwrap(_)
wave_eq1;
\end{cadabra}
\begin{dmath*}
  \mathbf{e}_{\mu \nu} \partial_{\sigma}\,^{\sigma}\left(\exp\left(i
      k_{\lambda} x^{\lambda}\right)\right)+\bar{\mathbf{e}}_{\mu \nu}
  \partial_{\sigma}\,^{\sigma}\left(\exp\left(-i k_{\lambda}
      x^{\lambda}\right)\right) = 0
\end{dmath*}
Note that
$\partial_\sigma^{~\sigma} = \partial_\sigma \partial^\sigma$. In
order to calculate the derivative $\partial^{\sigma}$ we use
substitute in \verb|rule_1| from above and use the
\verb|replace_index| textttrithm imported at the beginning to ensure
that all the dummy pairs have unique names
\begin{cadabra}
substitute(_, rule_1)
replace_index(_, r'\exp', r'\lambda', r'\psi');
\end{cadabra}
\begin{dmath*}
  \mathbf{e}_{\mu \nu} \partial_{\sigma}\left(\exp\left(i k_{\psi}
      x^{\psi}\right) \partial^{\sigma}\left(i k_{\lambda}
      x^{\lambda}\right)\right)-\bar{\mathbf{e}}_{\mu \nu}
  \partial_{\sigma}\left(\exp\left(-i k_{\psi} x^{\psi}\right)
    \partial^{\sigma}\left(i k_{\lambda} x^{\lambda}\right)\right) = 0
\end{dmath*}
We now use the \verb|unwrap| textttrithm to take the constants out of
the derivative, substitute in the metric where there are derivatives
of the coordinate (i.e.
$\partial^{\mu}{x^{\alpha}} \rightarrow \eta^{\mu\alpha}$) and then
perform another \verb|unwrap|.
\begin{cadabra}
unwrap(_)
substitute(_, $\partial^{\mu}{x^{\alpha}} -> \eta^{\mu\alpha}$);
unwrap(_);
\end{cadabra}
\begin{dgroup*}
  \begin{dmath*}
    \mathbf{e}_{\mu \nu} i k_{\lambda}
    \partial_{\sigma}\left(\exp\left(i k_{\psi} x^{\psi}\right)
      \eta^{\sigma \lambda}\right)-\bar{\mathbf{e}}_{\mu \nu} i
    k_{\lambda} \partial_{\sigma}\left(\exp\left(-i k_{\psi}
        x^{\psi}\right) \eta^{\sigma \lambda}\right) = 0
  \end{dmath*}
  \begin{dmath*}
    \mathbf{e}_{\mu \nu} i k_{\lambda} \eta^{\sigma \lambda}
    \partial_{\sigma}\left(\exp\left(i k_{\psi}
        x^{\psi}\right)\right)-\bar{\mathbf{e}}_{\mu \nu} i k_{\lambda}
    \eta^{\sigma \lambda} \partial_{\sigma}\left(\exp\left(-i k_{\psi}
        x^{\psi}\right)\right) = 0
  \end{dmath*}
\end{dgroup*}
We now calculate the derivative of the exponentional using
\verb|rule_3| and do another dummy index substitution as above
\begin{cadabra}
substitute(_, rule_3)
replace_index(_, r'\exp', r'\psi', r'\pi');
\end{cadabra}
\begin{dmath*}
  \mathbf{e}_{\mu \nu} i k_{\lambda} \eta^{\sigma \lambda} \exp\left(i
    k_{\pi} x^{\pi}\right) \partial_{\sigma}\left(i k_{\psi}
    x^{\psi}\right)+\bar{\mathbf{e}}_{\mu \nu} i k_{\lambda}
  \eta^{\sigma \lambda} \exp\left(-i k_{\pi} x^{\pi}\right)
  \partial_{\sigma}\left(i k_{\psi} x^{\psi}\right) = 0
\end{dmath*}
To finish off this calculation we once more use \verb|unwrap| to take
out the constants before performing the substitution
$\partial_{\mu}{x^{\alpha}} \rightarrow \delta_{\mu}^{~\alpha}$ and
cleaning up the result
\begin{cadabra}
unwrap(_);
substitute(_, $\partial_{\mu}{x^{\alpha}} -> \delta_{\mu}^{\alpha}$) # Can also do \partial_{\mu}{x^{\alpha}}:KroneckerDelta
eliminate_kronecker(_);
eliminate_metric(_)
sort_product(_)
manip.multiply_through(wave_eq1, $-1$);
\end{cadabra}
\begin{dgroup*}[noalign]
  \begin{dmath*}
    -\mathbf{e}_{\mu \nu} k_{\lambda} \eta^{\sigma \lambda} \exp\left(i k_{\pi} x^{\pi}\right) k_{\psi} \partial_{\sigma}{x^{\psi}}-\bar{\mathbf{e}}_{\mu \nu} k_{\lambda} \eta^{\sigma \lambda} \exp\left(-i k_{\pi} x^{\pi}\right) k_{\psi} \partial_{\sigma}{x^{\psi}} = 0
  \end{dmath*}
  \begin{dmath*}
    -\mathbf{e}_{\mu \nu} k_{\lambda} \eta^{\psi \lambda} \exp\left(i k_{\pi} x^{\pi}\right) k_{\psi}-\bar{\mathbf{e}}_{\mu \nu} k_{\lambda} \eta^{\psi \lambda} \exp\left(-i k_{\pi} x^{\pi}\right) k_{\psi} = 0
  \end{dmath*}
  \begin{dmath*}
    -\exp\left(i k_{\pi} x^{\pi}\right) \mathbf{e}_{\mu \nu} k_{\lambda} k^{\lambda}-\exp\left(-i k_{\pi} x^{\pi}\right) \bar{\mathbf{e}}_{\mu \nu} k_{\lambda} k^{\lambda} = 0
  \end{dmath*}
  \begin{dmath*}
    \exp\left(i k_{\pi} x^{\pi}\right) \mathbf{e}_{\mu \nu} k_{\lambda} k^{\lambda}+\exp\left(-i k_{\pi} x^{\pi}\right) \bar{\mathbf{e}}_{\mu \nu} k_{\lambda} k^{\lambda} = 0
  \end{dmath*}
\end{dgroup*}
This equation must be true for every value of $\mathbf{e}_{\mu\nu}$
and $\bar{\mathbf{e}}_{\mu\nu}$. Considering for example
$\mathbf{e}_{\mu\nu}$, the first wave relation can be extracted using
the \verb|get_basis_component| 
\begin{cadabra}
wave_eq1 = get_basis_component(wave_eq1, $\pol1{#}\exp(A??)$);
\end{cadabra}
\begin{dmath*}
  \boxed{k_{\lambda} k^{\lambda} = 0} \qquad \mbox{(First Wave Relation)}
\end{dmath*}
Clearly, we would have achieved the same result using
\verb|\pol2{#}\exp(A??)| as the basis. 

\subsection{Second wave-relation}

As $h_{\mu\nu}$ is a small perturbation around the flat spacetime, its
indices can been raised or lowered using the Minkowski metric
$\eta_{\mu\nu}$
\begin{cadabra}
sol2 := @(sol);
manip.multiply_through(_, $\eta^{\gamma\mu}$)
distribute(_)
eliminate_metric(_);
\end{cadabra}
\begin{dgroup*}[noalign]
  \begin{dmath*}
    \p{1}{h}_{\mu \nu} = \mathbf{e}_{\mu \nu} \exp\left(i k_{\lambda} x^{\lambda}\right)+\bar{\mathbf{e}}_{\mu \nu} \exp\left(-i k_{\lambda} x^{\lambda}\right)
  \end{dmath*}
  \begin{dmath*}
    \p{1}{h}^{\gamma}\,_{\nu} = \mathbf{e}^{\gamma}\,_{\nu} \exp\left(i k_{\lambda} x^{\lambda}\right)+\bar{\mathbf{e}}^{\gamma}\,_{\nu} \exp\left(-i k_{\lambda} x^{\lambda}\right)
  \end{dmath*}
\end{dgroup*}
Let us impose that the plane-wave $h^\gamma_{~\nu}$ satisfies the
harmonic-gauge condition and use \verb|distribute| and \verb|unwrap|
to move the constants out of the derivative
\begin{cadabra}
wave_eq2 := @(gauge_eq);
substitute(_, sol2)
distribute(_);
\end{cadabra}
\begin{dgroup*}[noalign]
  \begin{dmath*}
    \partial_{\mu}{\p{1}{h}^{\mu}\,_{\nu}} - \tfrac{1}{2}\partial_{\nu}{\p{1}{h}^{\mu}\,_{\mu}} = 0
  \end{dmath*}
  \begin{dmath*}
    \partial_{\mu}\left(\mathbf{e}^{\mu}\,_{\nu} \exp\left(i k_{\lambda} x^{\lambda}\right)\right)+\partial_{\mu}\left(\bar{\mathbf{e}}^{\mu}\,_{\nu} \exp\left(-i k_{\lambda} x^{\lambda}\right)\right) - \tfrac{1}{2}\partial_{\nu}\left(\mathbf{e}^{\mu}\,_{\mu} \exp\left(i k_{\lambda} x^{\lambda}\right)\right) - \tfrac{1}{2}\partial_{\nu}\left(\bar{\mathbf{e}}^{\mu}\,_{\mu} \exp\left(-i k_{\lambda} x^{\lambda}\right)\right) = 0
  \end{dmath*}
\end{dgroup*}
As with the the first wave-relation, we use \verb|unwrap| to move the
constants outside of the derivative, substitute in \verb|rule_3| to
calculate the derivative before cleaning up the indices with
\verb|replace_index| and calling \verb|unwrap| again
\begin{cadabra}
unwrap(_)
substitute(_, rule_3)
replace_index(_, r'\exp', r'\lambda', r'\psi')
unwrap(_);
\end{cadabra}
\begin{dmath*}
  \mathbf{e}^{\mu}\,_{\nu} \exp\left(i k_{\psi} x^{\psi}\right) i k_{\lambda} \partial_{\mu}{x^{\lambda}}-\bar{\mathbf{e}}^{\mu}\,_{\nu} \exp\left(-i k_{\psi} x^{\psi}\right) i k_{\lambda} \partial_{\mu}{x^{\lambda}} - \tfrac{1}{2}\mathbf{e}^{\mu}\,_{\mu} \exp\left(i k_{\psi} x^{\psi}\right) i k_{\lambda} \partial_{\nu}{x^{\lambda}}+\tfrac{1}{2}\bar{\mathbf{e}}^{\mu}\,_{\mu} \exp\left(-i k_{\psi} x^{\psi}\right) i k_{\lambda} \partial_{\nu}{x^{\lambda}} = 0
\end{dmath*}
As before we replace $\partial_{\mu}{x^{\lambda}}$ with the Kronecker
delta which we then resolve with \verb|eliminate_kronecker|. We then
separate out the coefficients of the two different exponential
functions using \verb|factor_out|
\begin{cadabra}
substitute(wave_eq2, $\partial_{\mu}{x^{\lambda}} -> \delta_{\mu}^{\lambda}$);
eliminate_kronecker(wave_eq2);
factor_out(wave_eq2,$\exp(i k_{\psi} x^{\psi}), \exp(-i k_{\psi} x^{\psi})$);
\end{cadabra}
\begin{dgroup*}[noalign]
  \begin{dmath*}
    \mathbf{e}^{\mu}\,_{\nu} \exp\left(i k_{\psi} x^{\psi}\right) i k_{\lambda} \delta_{\mu}\,^{\lambda}-\bar{\mathbf{e}}^{\mu}\,_{\nu} \exp\left(-i k_{\psi} x^{\psi}\right) i k_{\lambda} \delta_{\mu}\,^{\lambda} - \tfrac{1}{2}\mathbf{e}^{\mu}\,_{\mu} \exp\left(i k_{\psi} x^{\psi}\right) i k_{\lambda} \delta_{\nu}\,^{\lambda}+\tfrac{1}{2}\bar{\mathbf{e}}^{\mu}\,_{\mu} \exp\left(-i k_{\psi} x^{\psi}\right) i k_{\lambda} \delta_{\nu}\,^{\lambda} = 0
  \end{dmath*}
  \begin{dmath*}
    \mathbf{e}^{\lambda}\,_{\nu} \exp\left(i k_{\psi} x^{\psi}\right) i k_{\lambda}-\bar{\mathbf{e}}^{\lambda}\,_{\nu} \exp\left(-i k_{\psi} x^{\psi}\right) i k_{\lambda} - \tfrac{1}{2}\mathbf{e}^{\mu}\,_{\mu} \exp\left(i k_{\psi} x^{\psi}\right) i k_{\nu}+\tfrac{1}{2}\bar{\mathbf{e}}^{\mu}\,_{\mu} \exp\left(-i k_{\psi} x^{\psi}\right) i k_{\nu} = 0
  \end{dmath*}
  \begin{dmath*}
    \exp\left(i k_{\psi} x^{\psi}\right) \left(\mathbf{e}^{\lambda}\,_{\nu} i k_{\lambda} - \tfrac{1}{2}\mathbf{e}^{\mu}\,_{\mu} i k_{\nu}\right)+\exp\left(-i k_{\psi} x^{\psi}\right) \left(-\bar{\mathbf{e}}^{\lambda}\,_{\nu} i k_{\lambda}+\tfrac{1}{2}\bar{\mathbf{e}}^{\mu}\,_{\mu} i k_{\nu}\right) = 0
  \end{dmath*}
\end{dgroup*}
This equation must be true for every value of $\mathbf{e}_{\mu\nu}$
and $\bar{\mathbf{e}}_{\mu\nu}$. Considering for example
$\mathbf{e}_{\mu\nu}$, the second wave relation can be extracted using
the \texttt{get_basis_component} function.
\begin{cadabra}
wave_eq2 = get_basis_component(wave_eq2,$\exp(i k_{\psi} x^{\psi})$);
\end{cadabra}
\begin{equation}
  \boxed{\mathbf{e}^{\lambda}\,_{\nu} i k_{\lambda} -
    \tfrac{1}{2}\mathbf{e}^{\mu}\,_{\mu} i k_{\nu} = 0} \qquad \mbox{(Second Wave Relation)}
\end{equation}


\section{Higher-order gravitational-waves solutions in vacuum}
\label{sec:highgw}

In the previous Sections \ref{sec:fistgw} and \ref{sec:gwrel} we only
exhibited first-order and second-order perturbative tensor examples.
It is clear that the tools developed in Section \ref{sec:tenspert} can
be also managed to deal with higher-order problems. For reasons of
clarity, we now apply our machinery to reproduce some results obtained
in Ref.  \cite{Arcos:2015uqa}, where a complete analytical analysis of
vacuum high-order gravitational waves solutions is given. The scope of
the following Section \ref{sec:highgw} is to provide a computational
counterpart of such analysis, highlighting the natural predisposition
of \verb+Cadabra+ Software to treat the heavy and onerous nature of
high-order tensor calculations. The formalism is introduced in Section
\ref{sec:tensPertGR}. 

Following the same method exhibited in Ref. \cite{Arcos:2015uqa} (see
Section 2.6, Eqs.~(63) and~(64) of the reference paper), let us
consider a general parametrization for background, first-order, and
higher-order gravitational wave solutions that travel in the
$z$-coordinate direction
\begin{gather}
  \label{eq:parsolh01}
  \p{0}{h}_{\mu\nu}=0
  \qquad
  \p{1}{h}_{\mu\nu}=\eta_{\mu\alpha}A^{\alpha}\!_{\nu}\exp\left(ik_{\rho}x^{\rho}\right)\,,
  \\
  \label{eq:parsolh}
  \p{2n}{h_{\mu\nu}} = \Phi^n \left(\p{2n}{a}\eta_{\mu\alpha}\alpha^{\alpha}\!_{\nu}
    + i z \omega c^{-1}\p{2n}{b}\eta_{\mu\alpha}\beta^{\alpha}\!_{\nu}\right)\exp\left(i2nk_{\rho}x^{\rho}\right)\,,
\end{gather}
where $2n$ is the perturbative order, with $n = 1, 2, 3, \ldots, N$.
Clearly, odd-order solutions are null. In definition
\eqref{eq:parsolh} we introduced the following quantities 
\begin{gather}
  \label{eq:coeff}
  \Phi\equiv \p{1}{A}^{\mu}\,_{\nu} \p{1}{A}^{\nu}\,_{\mu} = \p{1}{A}^{\mu}\,_{\nu} \p{1}{A}_{\mu}\,^{\nu}\,,
  \\
  \label{eq:matr}
  \alpha^{\mu}\!_{\nu}=\delta^{\mu}\!_{\nu}
  \equiv
  \begin{pmatrix} 
    1 & 0 & 0&0\\
    0 & 1 & 0&0\\
    0 & 0 & 1&0\\
    0 & 0 & 0&1
  \end{pmatrix}
  \qquad
  \beta^{\mu}\!_{\nu}
  \equiv
  \begin{pmatrix} 
    1 & 0 & 0&-1\\
    0 & 0 & 0&0\\
    0 & 0 & 0&0\\
    1 & 0 & 0&-1
  \end{pmatrix}
  \qquad
  k^{\mu}
  \equiv
  \begin{pmatrix} 
    \omega/c \\
    0 \\
    0 \\
    \omega/c
  \end{pmatrix}
  ,
\end{gather}
with $\p{1}{A}^{\mu}\,_{\nu}$ the symmetric first-order polarization
tensor expressed in terms of cross ($\p{\times}{h}$) and plus
($\p{+}{h}$) states 
\begin{equation}
  \label{eq:Apol}
  \p{1}{A}^{\mu}\!_{\nu}
  \equiv
  \begin{pmatrix} 
      \p{+}{h} & \p{\times}{h} & 0&0\\
      \p{\times}{h}  & -\p{+}{h} & 0&0\\
      0 & 0 & 0&0\\
      0 & 0 & 0&0
   \end{pmatrix}.
\end{equation}
In Eq. \eqref{eq:parsolh} the numerical coefficients
$\{\p{2n}{a},\p{2n}{b}\}$ will be the result of our computations, and
must be found using Einstein's equations in vacuum and the harmonic
coordinate conditions at each perturbative order. 
\begin{equation}
  \label{eq:eqs}
  R_{\mu\nu}=0 \qquad g^{\mu\nu} \Gamma^{\sigma}_{\mu\nu} =0 ,
\end{equation}
Due to gravitational wave relations, at each perturbative order the
following gauge conditions are satisfied: 
\begin{equation}
  \label{eq:GWsRel}
  k^{\mu} k_{\mu} = 0
  \qquad
  k^{\mu} \p{1}{A}_{\mu\nu} = 0
  \qquad
  k_{\mu} \beta^{\mu\nu} = 0
\end{equation}

\subsection{Definitions and modules}
\label{sec:demhigh}

As with the notebooks from previous sections we begin by importing the
results of Section~\ref{sec:tenspert}~and functionality from the
Cadabra standard library. 

\StartLineAt{1}
\begin{cadabra}
import libraries.config as config
config.maxPertOrder=4
from sec4_TensorPerturbationsGR import *
from cdb.utils.indices import replace_index
\end{cadabra}

The \verb+header.cnb+ module has been automatically imported, whereas
\verb+sec4_TensorPerturbationsGR+ imports our dictionaries
\verb+gUpp+, \verb+connection+, \verb+ricciTensor+, etc. containing
the perturbative decompositions of the main tensorial quantities we
will employ. The maximum perturbative order we will deal with is fixed
by the \verb+maxPertOrder+ variable, which is valorized into the
\verb+sec4_TensorPerturbationsGR+ module. This value must be set
before executing the calculations. We suggest to use as a starting
point low perturbative orders to do not face honerous computational
times during the first tests. In our case, we set the value
\verb+maxPertOrder=10+.

\subsection{Objects and components}
\label{sec.comp}

The scope of this section is to reproduce the method obtained in
Ref.~\citep[sec.~2.6]{Arcos:2015uqa}, in particular the parameterised
solutions presented in Eqs.~\eqref{eq:parsolh01}
and~\eqref{eq:parsolh} (with related tensorial parameters), using a
\emph{translation} into \verb+Cadabra+ objects. We start making the following
property and equation definitions
\ContinueLineNumber
\begin{cadabra}
# Normalization factor
n1 := A^{\mu}_{\nu} A^{\nu}_{\mu} = \Phi.
n2 := A^{\mu}_{\nu} A_{\mu}^{\nu} = \Phi.
# Properties of parametrizing objects
A_{\mu\nu}::Symmetric.
B_{\mu\nu}::Symmetric.
\exp{#}::Depends(\partial{#}).
Z::Depends(\partial{#}).
for i in range(maxPertOrder+1):
  LaTeXForm(Ex(f"a{i}"), Ex(r'"\,^{^{(
  LaTeXForm(Ex(f"b{i}"), Ex(r'"\,^{^{(
\end{cadabra}
Note that we use $Z$ in place of the coordinate $z$ in order to avoid
the \verb|sympy_bridge| problem (as \verb+Cadabra+ can not currently export
expression which use the \verb|Depends| property to Sympy) when we
\verb|evaluate()| the expressions. We do the algebraic manipulations
using Z and then make the substitution $Z \rightarrow z$ when doing
evaluations.

We now construct a list \verb|sol| with parameterized solutions up to
\verb|maxPertOrder|. In order to avoid introduce extra
metric contractions later, we define these for all the necessary positions 
of indices using the auxiliary function
\verb|all_index_positions|:
\begin{cadabra}
def all_index_positions(ex):
  tmp := @(ex).
  res = Ex(r"\comma")
  res.top().append_child($@(ex)$)
  for contraction, index in [($\eta^{\mu \lambda}$, r"\mu"), ($\eta^{\lambda \nu}$, r"\nu"), ($\eta_{\mu \lambda}$, r"\mu")]:
    manip.multiply_through(ex, contraction)
    distribute(ex)
    eliminate_metric(ex)
    eliminate_kronecker(ex)
    replace_index(ex, r"\equals", r"\lambda", index)
    res.top().append_child($@(ex)$)
  return res
\end{cadabra}

\begin{cadabra}
sol = []
# Zero and First order by hand
sol.append($h0^{\mu\nu} = 0$)
sol.append($h1_{\mu\nu} = \eta_{\mu\alpha}A^{\alpha}_{\nu}\exp(i k_{\rho} x^{\rho})$)
# Higher order solutions
for i in range(2, maxPertOrder+1):
  # Construct symbols based on current pert order
  iex, ai, bi, hi = Ex(i), Ex(f'a{i}'), Ex(f'b{i}'), Ex(f'h{i}')
  # Define solutions for each order and append to `sol`
  if i
    # Define the solution for even orders
    ex1 := @(ai) \eta_{\mu\alpha} \delta^{\alpha}_{\nu}.
    ex2 := i @(bi) Z \omega/c \eta_{\mu\alpha} B^{\alpha}_{\nu}.
    ex3 := \exp(@(iex) i k_{\rho}x^{\rho}).
    ex := @(hi)_{\mu\nu} = \Phi**{@(iex)/2} (@(ex1) + @(ex2)) @(ex3).
    # Append the solution to the sol list
    expand_power(ex)
    sol.append(ex)  
  else:
    # Odd orders are all zero
    sol.append($@(hi)_{\mu \nu} = 0$)
\end{cadabra}
We will print out a couple of the elements in this list to ensure that
the construction is 
correct
\begin{cadabra}[numbers=none]
sol[-2][0];
sol[-1][0];
\end{cadabra}
\begin{dgroup*}[noalign]
  \begin{dmath*}
    \p{3}{h}_{\mu \nu} = 0
  \end{dmath*}
  \begin{dmath*}
    \p{4}{h}_{\mu \nu} = \Phi \Phi \Big(\p{4}{a} \eta_{\mu \alpha} \delta^{\alpha}\,_{\nu}+i \p{4}{b} Z \omega {c}^{-1} \eta_{\mu \alpha} B^{\alpha}\,_{\nu}\Big) \exp\Big(4i k_{\rho} x^{\rho}\Big)
  \end{dmath*}
\end{dgroup*}

At each perturbative order the gauge conditions $k^\mu k_\mu = 0$,
$k^\mu A_{\mu\nu} = 0$ and $k_\mu B^{\mu\nu} = 0$ exposed in
\eqref{eq:GWsRel} are satisfied which we express for all index
positions here:
\StartLineAt{49}
\begin{cadabra}
# Create substitution rules from the gauge parameters
gau1 := k^{\mu} k_{\mu} -> 0       , k_{\mu} k^{\mu} -> 0.
gau2 := k^{\mu} A_{\mu\nu} -> 0    , k^{\mu} A_{\nu\mu} -> 0,
        k^{\mu} A_{\mu}^{\nu} -> 0 , k^{\mu} A^{\nu}_{\mu} -> 0,
        k_{\mu} A^{\mu\nu} -> 0    , k_{\mu} A^{\nu\mu} -> 0,
        k_{\mu} A^{\mu}_{\nu} -> 0 , k_{\mu} A_{\nu}^{\mu} -> 0.
gau3 := k^{\mu} B_{\mu\nu} -> 0    , k^{\mu} B_{\nu\mu} -> 0,
        k^{\mu} B_{\mu}^{\nu} -> 0 , k^{\mu} B^{\nu}_{\mu} -> 0,
        k_{\mu} B^{\mu\nu} -> 0    , k_{\mu} B^{\nu\mu} -> 0,
        k_{\mu} B^{\mu}_{\nu} -> 0 , k_{\mu} B_{\nu}^{\mu} -> 0.
\end{cadabra}

\subsection{Evaluation of the parameters}
\label{sec:pareval}

With reference to \cite{Arcos:2015uqa}, we will now calculate the
components of the parameters $k_{\mu}$, $A_{\mu\nu}$ and $B_{\mu\nu}$. 

We first define a function \verb|evaluate_and_complete| which
evaluates the components of an equation and uses these to complete a
set of substitution rules.
\ContinueLineNumber
\begin{cadabra}
def evaluate_and_complete(ex, rule, other_components):
  tmp = evaluate(rule, join(ex, other_components), rhsonly=True)
  subrules = comp.components_to_subrule(tmp)
  for subrule in subrules[r'\arrow']:
    subrule.name = r'\equals'
    ex.top().append_child(subrule)
  return ex
\end{cadabra}
After definining the components of the
Minkowski metric we then use this to write out the components of the
parameters $k_{\mu}$, $A_{\mu \nu}$, $B_{\mu \nu}$ and \(\Phi\),
\begin{cadabra}
# Minkowski flat metric
mink := { \eta_{t t} = 1, \eta_{x x} = -1, \eta_{y y} = -1, \eta_{z z} = -1 }.
complete(mink, $\eta^{\mu\nu}$);
# Position 
x := {x_{t} = t, x_{x} = -x, x_{y} = -y, x_{z} = -z};
# Wavenumber (by Eq. 24)
k := {k^{t} = \omega/c, k^{z} = \omega/c}.
evaluate_and_complete(k, $k_{\mu} = \eta_{\mu\rho}k^{\rho}$, mink);
# Solution's parameters (from eqns (65) and (25))
A := {A^{x}_{x}=-p, A^{x}_{y}=-m, A^{y}_{x}=-m, A^{y}_{y}=p}.
evaluate_and_complete(A, $A^{\mu\nu} = \eta^{\nu\rho}A^{\mu}_{\rho}$, mink)
evaluate_and_complete(A, $A_{\mu\nu} = \eta_{\mu\rho}A^{\rho}_{\nu}$, mink)
evaluate_and_complete(A, $A_{\mu}^{\nu} = \eta^{\nu\sigma}A_{\mu\sigma}$, mink);
B := {B^{t}_{t}=1,B^{t}_{z}=-1,B^{z}_{t}=1,B^{z}_{z}=-1}.
evaluate_and_complete(B, $B^{\mu\nu} = \eta^{\nu\rho}B^{\mu}_{\rho}$, mink)
evaluate_and_complete(B, $B_{\mu\nu} = \eta_{\mu\rho}B^{\rho}_{\nu}$, mink)
evaluate_and_complete(B, $B_{\mu}^{\nu} = \eta^{\nu\sigma}B_{\mu\sigma}$, mink);
# Normalisation
norm = manip.swap_sides($@(n1)$)
evaluate(norm, A);
\end{cadabra}
to get the following evaluated components
\begin{dgroup*}[noalign]
  \begin{dmath*}
    {\Big[\eta_{t t} = 1, \eta_{x x} = -1, \eta_{y y} = -1, \eta_{z z} = -1, \eta^{t t} = 1, \eta^{x x} = -1, \eta^{y y} = -1, \eta^{z z} = -1\Big]}
  \end{dmath*}
  \begin{dmath*}
    {\Big[x_{t} = t, x_{x} = -x, x_{y} = -y, x_{z} = -z\Big]}
  \end{dmath*}
  \begin{dmath*}
    {\Big[k^{t} = \omega {c}^{-1}, k^{z} = \omega {c}^{-1}, k_{t} = \omega {c}^{-1}, k_{z} = -\omega {c}^{-1}\Big]}
  \end{dmath*}
  \begin{dmath*}
    \Big[{A^{x}\,_{x} = -p, A^{x}\,_{y} = -m, A^{y}\,_{x} = -m, A^{y}\,_{y} = p,}~\discretionary{}{}{} {A^{x x} = p, A^{x y} = m, A^{y x} = m, A^{y y} = -p,}~\discretionary{}{}{} {A_{x x} = p, A_{x y} = m, A_{y x} = m, A_{y y} = -p,}~\discretionary{}{}{} {A_{x}\,^{x} = -p, A_{x}\,^{y} = -m, A_{y}\,^{x} = -m, A_{y}\,^{y} = p}\Big]
  \end{dmath*}
  \begin{dmath*}
    \Big[{B^{t}\,_{t} = 1, B^{t}\,_{z} = -1, B^{z}\,_{t} = 1, B^{z}\,_{z} = -1,}~\discretionary{}{}{} {B^{t t} = 1, B^{t z} = 1, B^{z t} = 1, B^{z z} = 1,}~\discretionary{}{}{} {B_{t t} = 1, B_{t z} = -1, B_{z t} = -1, B_{z z} = 1,}~\discretionary{}{}{} {B_{t}\,^{t} = 1, B_{t}\,^{z} = 1, B_{z}\,^{t} = -1, B_{z}\,^{z} = -1}\Big]
  \end{dmath*}
  \begin{dmath*}
    \Phi = 2{m}^{2}+2{p}^{2}
  \end{dmath*}
\end{dgroup*}

\subsection{Einstein Equation and Harmonic Gauge}

To evaluate the numerical values of the parameters $\p{i}{a}$ and
$\p{i}{b}$ we will need to calucalate the components of the Einstein
equations and harmonic gauge condition at different perturbative
orders. We begin by calculating the symbolic decomposition of these
equations with \verb|subsPertSums| function defined in Subsection
\ref{sec:func1}
\begin{cadabra}
# Einstein Equation: symbolic decomposition
ein := 0 = rc_{\mu\nu};
ein = subsPertSums(ein,pertLabel,maxPertOrder,ricciTensor['sym']);
# Harmonic Gauge condition: symbolic decomposition
gau := 0 = g^{\mu\nu} ch^{\sigma}_{\mu\nu};
gau = subsPertSums(gau,pertLabel,maxPertOrder,gUpp['sym'],connection['sym']);
\end{cadabra}
\begin{dgroup*}[noalign]
  \begin{dmath*}
    0 = R_{\mu \nu}
  \end{dmath*}
  \begin{dmath*}
    0 = \p{0}{R}_{\mu \nu}+\p{1}{R}_{\mu \nu}+\p{2}{R}_{\mu \nu}+\p{3}{R}_{\mu \nu}+\p{4}{R}_{\mu \nu}
  \end{dmath*}
  \begin{dmath*}
    0 = g^{\mu \nu} \Gamma^{\sigma}\,_{\mu \nu}
  \end{dmath*}
  \begin{dmath*}
    0 = \p{0}{g}^{\mu \nu} \p{0}{\Gamma}^{\sigma}\,_{\mu \nu}+\p{0}{g}^{\mu \nu} \p{1}{\Gamma}^{\sigma}\,_{\mu \nu}+\p{0}{g}^{\mu \nu} \p{2}{\Gamma}^{\sigma}\,_{\mu \nu}+\p{0}{g}^{\mu \nu} \p{3}{\Gamma}^{\sigma}\,_{\mu \nu}+\p{0}{g}^{\mu \nu} \p{4}{\Gamma}^{\sigma}\,_{\mu \nu}+\p{1}{g}^{\mu \nu} \p{0}{\Gamma}^{\sigma}\,_{\mu \nu}+\p{1}{g}^{\mu \nu} \p{1}{\Gamma}^{\sigma}\,_{\mu \nu}+\p{1}{g}^{\mu \nu} \p{2}{\Gamma}^{\sigma}\,_{\mu \nu}+\p{1}{g}^{\mu \nu} \p{3}{\Gamma}^{\sigma}\,_{\mu \nu}+\p{2}{g}^{\mu \nu} \p{0}{\Gamma}^{\sigma}\,_{\mu \nu}+\p{2}{g}^{\mu \nu} \p{1}{\Gamma}^{\sigma}\,_{\mu \nu}+\p{2}{g}^{\mu \nu} \p{2}{\Gamma}^{\sigma}\,_{\mu \nu}+\p{3}{g}^{\mu \nu} \p{0}{\Gamma}^{\sigma}\,_{\mu \nu}+\p{3}{g}^{\mu \nu} \p{1}{\Gamma}^{\sigma}\,_{\mu \nu}+\p{4}{g}^{\mu \nu} \p{0}{\Gamma}^{\sigma}\,_{\mu \nu}
  \end{dmath*}
\end{dgroup*}
We now define the function \verb|subsParamSol| which takes a symbolic
decomposition \verb|ex| of order \verb|pertOrder| and substitutes in
the parameterizations in the \verb|sol| list.

To begin, we loop over all the metric perturbations $\p{i}{h}$ up to
\verb|maxPertOrder| and substitute in the parameterized solutions from \verb|sol|
\begin{cadabra}
def subsParamSol(ex,pertOrder):
  # Substitute in the parameterized solutions
  for j in range(1,pertOrder+1): 
    substitute(ex, sol[j])
\end{cadabra}
We then unnest the resulting expression so that derivatives of
products of exponentials are replaced with products of derivatives.   
\begin{cadabra}  
  # Get derivatives of \exp only. Maximum 2nd derivatives in expression
  converge(ex):
    distribute(ex)
    product_rule(ex)
    unwrap(ex)
\end{cadabra}
We also need to lower the indices of the derivatives,
\begin{cadabra}  
  # Calculate derivatives of \exp: (a) lower indices on derivatives
  substitute(ex, $\partial^{\mu}{Q??} -> \eta^{\mu\nu}\partial_{\nu}{Q??}$)
  substitute(ex, $\partial^{\mu}_{\nu}{Q??} -> \eta^{\mu\rho}\partial_{\rho\nu}{Q??}$)
  substitute(ex, $\partial_{\mu}^{\nu}{Q??} -> \eta^{\nu\rho}\partial_{\mu\rho}{Q??}$)
\end{cadabra}  
We are now ready to perform the differentiation by the substituting in
the definitions $\partial_{\nu}{\exp(a \, i \, k_{\rho} x^{\rho})} \to a \, i \,
k_{\nu}$ and $\partial_{\mu \nu}{\exp(a \, i \, k_{\rho} x^{\rho})} \to - a^2 \,
k_{\mu} k_{\nu}$,
\begin{cadabra}  
  # Calculate derivatives of \exp: (b) execute derivatives with substitutions
  for j in range(1,pertOrder+1):
    jex = Ex(j)
    substitute(ex, $\partial_{\nu}{\exp(@(jex) i k_{\rho}x^{\rho})} -> @(jex) i k_{\nu}$)
    substitute(ex, $\partial_{\mu\nu}{\exp( @(jex) i k_{\rho}x^{\rho})} -> -@(jex) @(jex) k_{\mu}k_{\nu}$)   
    substitute(ex, $\exp(@(jex) i k_{\rho}x^{\rho}) -> 1$)
\end{cadabra}
Now that we have substituted in these expressions we can insert the
gauge symmetries
\begin{cadabra}
  # Insert gauge symmetries
  substitute(ex,gau1)
  substitute(ex,gau2)
  substitute(ex,gau3)
\end{cadabra}
To finish off, we perform some final simplifications by setting the second 
derivatives of the coordinates to \(0\), removing trace terms of \(B\) and
eliminating any contracted metrics and Kronecker deltas:
\begin{cadabra}
  eliminate_metric(ex, repeat=True)
  eliminate_kronecker(ex, repeat=True)
  # Set second derivatives of coordinates to 0
  substitute(ex, $\partial_{\mu\nu}{Z} -> 0$)
  substitute(ex, $\partial_{\mu}^{\nu}{Z} -> 0$)
  substitute(ex, $\partial^{\mu}_{\nu}{Z} -> 0$)
  substitute(ex, $\partial^{\mu\nu}{Z} -> 0$)
  # Trace of B matrix is zero
  substitute(ex, $B^{\rho}_{\rho} -> 0$)
  substitute(ex, $B_{\rho}^{\rho} -> 0$)
  return ex
\end{cadabra}

We use this to substitute in the parameterized solutions to
$g^{\mu \nu}$, $\Gamma^{\rho}_{~ \mu \nu}$ and $R_{\mu \nu}$ which
will be needed to fully expand the solutions
\begin{cadabra}
# Copy the the lists calculated in section 4
evalG = gUpp['ord'].copy()
evalC = connection['ord'].copy()
evalR = ricciTensor['ord'].copy()
# Substitute in the orders of the parameterized solutions
for i in range(1,maxPertOrder+1):
  evalG[i] = subsParamSol(evalG[i],i)
  evalC[i] = subsParamSol(evalC[i],i)
  evalR[i] = subsParamSol(evalR[i],i)
\end{cadabra}
We define a final function \verb|evalParam| which evaluates \verb|ex|
over the given components and simplifies the result
\begin{cadabra}
def evalParam(ex, components):
  ex = comp.get_component(ex, components)
  # Lower indices
  substitute(ex, $Z->z$)
  distribute(ex)
  eliminate_kronecker(ex, repeat=True)
  eliminate_metric(ex, repeat=True)
  # Extract component and enter values
  expand_dummies(ex, join(A, join(B, join(k, mink))))
  substitute(ex, join(k, mink))
  substitute(ex, $\partial_{z}{z} -> 1, \partial_{t}{z} -> 0, \partial_{x}{z} -> 0, \partial_{y}{z} -> 0$)
  substitute(ex, $\partial^{z}{z} -> -1, \partial^{t}{z} -> 0, \partial^{x}{z} -> 0, \partial^{y}{z} -> 0$)
  substitute(ex, norm)
  # Clean up
  collect_factors(ex)
  distribute(ex)
  eliminate_kronecker(ex)
  return ex
\end{cadabra}

\subsection{Calculation of coefficients}
\label{sec:coeffcalc}

We are now ready to calculate the coefficients $a$ and $b$. We will
keep them in two lists in a dictionary such that the $i$th element
contains the value of $\p{i}{a}$ or $\p{i}{b}$ We define and
initialise these here with the first two orders which are trivial:
\begin{cadabra}
# Define the list of coefficients
coeffs = {
  'a': [None] * (maxPertOrder+1),
  'b': [None] * (maxPertOrder+1)
}
coeffs['a'][0], coeffs['b'][0] = $a0 -> 0$, $b0 -> 0$
coeffs['a'][1], coeffs['b'][1] = $a1 -> 0$, $b1 -> 0$
\end{cadabra}

We now iterate through the remaining orders calculating the values of
these coefficients at these orders by examining the Einstein and gauge
equations at this order and substituting in the values calculated for
the lower orders.

We begin by defining expressions for the variables $\p{i}{a}$ and
$\p{i}{b}$ so that we can pull them in to our equations later,
\begin{cadabra}
for i in range(2, maxPertOrder+1):
  ai, bi = Ex(f'a{i}'), Ex(f'b{i}')
\end{cadabra}
For odd orders the values are always $0$ and so we make a special case
for these values,
\begin{cadabra}
  # Check if the order is odd (and thus = 0) or even (and requires calculation)
  if i
    coeffs['a'][i] = $@(ai) -> 0$
    coeffs['b'][i] = $@(bi) -> 0$
\end{cadabra}
For even orders we begin by getting the \(i\)-th order of the equation
and substituting in the decompositions of the Einstein equation and
gauge conditions which we calculated earlier
\begin{cadabra}
  else:
    # Get the perturbative order of each equation
    eq1 = getEquationPertOrder(ein,pertLabel,i)
    eq2 = getEquationPertOrder(gau,pertLabel,i)
    # Substitute in the perturbed orders
    for j in range(i+1):
      substitute(eq1, evalR[j]) #ein
      substitute(eq2, evalG[j]) #gau
      substitute(eq2, evalC[j]) #gau
\end{cadabra}
We now simplify the gauge equation in the usual style by eliminating
metric tensors which can raise and lower indices, substituting in the
gauge conditions and enforcing the tracelessness of $B$,
\begin{cadabra}
    eliminate_metric(eq2, repeat=True)
    eliminate_kronecker(eq2, repeat=True)
    substitute(eq2, gau1)
    substitute(eq2, gau2)
    substitute(eq2, gau3)
    substitute(eq2, $B^{\rho}_{\rho}->0$)
    substitute(eq2, $B_{\rho}^{\rho}->0$)
\end{cadabra}
Then, we go through the previous perturbative orders which we have
already calculated numerical values for and substitute these in,
\begin{cadabra}
    # Substitute in the previously calculated coefficients
    for j in range(1,j):
      substitute(eq1, coeffs['a'][j])
      substitute(eq1, coeffs['b'][j])
      substitute(eq2, coeffs['a'][j])
      substitute(eq2, coeffs['b'][j])
\end{cadabra}
In order to calculate the numerical value at this order, we extract
the $t$ components of the equations and use the \verb|linsolve|
algorithm from the \verb|cdb.core.solve| library to solve for
$\p{i}{a}$ and $\p{i}{b}$ which we then append to the \verb|coeffs|
arrays
\begin{cadabra}
    # Evaluate equations
    eq1 = evalParam(eq1, $t, t$)
    eq2 = evalParam(eq2, $t$)
    # Solve the component equations as a linear system
    sols = solv.linsolve($@(eq1), @(eq2)$, $@(ai), @(bi)$)
    # Append our newly calculated solutions
    coeffs['a'][i] = sols[0][0]
    coeffs['b'][i] = sols[0][1]
\end{cadabra}
  
We can now display the numerical values of $\p{i}{a}$ and $\p{i}{b}$: 
\begin{cadabra}
for a, b in zip(*coeffs.values()):
  $@(a), @(b)$;
\end{cadabra}
\begin{dmath*}
{\Bigg[\p{1}{a} \rightarrow 0, \p{1}{b} \rightarrow 0\Bigg]} 
\quad~\discretionary{}{}{}
{\Bigg[\p{2}{a} \rightarrow \frac{3}{16}, \p{2}{b} \rightarrow  - \frac{1}{8}\Bigg]}
\quad~\discretionary{}{}{}
{\Bigg[\p{3}{a} \rightarrow 0, \p{3}{b} \rightarrow 0\Bigg]}
\quad~\discretionary{}{}{}
{\Bigg[\p{4}{a} \rightarrow  - \frac{1}{64}, \p{4}{b} \rightarrow  - \frac{1}{32}\Bigg]}
\end{dmath*}

Although exemplified the code for \verb+maxPertOrder=4+, we manage to
calculate the terms up to \verb+maxPertOrder=10+.
With respect to the reference paper, we have found from our
calculations that the $\p{4}{a}$ coefficient is $-\frac{1}{64}$
instead of $\frac{1}{64}$. As the use of this value has led to
calculating higher order coefficient which agree with the reference
paper, we believe that this is a typographical error in the paper. For
the sake of completeness, we report the remaining higher order
coefficients,
\begin{dmath*}
\discretionary{}{}{}
{\Bigg[\p{5}{a} \rightarrow 0, \p{5}{b} \rightarrow 0\Bigg]}
\quad~\discretionary{}{}{}
{\Bigg[\p{6}{a} \rightarrow \frac{13}{4608}, \p{6}{b} \rightarrow - \frac{1}{1536}\Bigg]}
\quad~\discretionary{}{}{}
{\Bigg[\p{7}{a} \rightarrow 0, \p{7}{b} \rightarrow 0\Bigg]}
\quad~\discretionary{}{}{}
{\Bigg[\p{8}{a} \rightarrow  - \frac{83}{131072}, \p{8}{b} \rightarrow  - \frac{65}{147456}\Bigg]}
\quad~\discretionary{}{}{}
{\Bigg[\p{9}{a} \rightarrow 0, \p{9}{b} \rightarrow 0\Bigg]}
\quad~\discretionary{}{}{}
{\Bigg[\p{10}{a} \rightarrow \frac{5023}{31457280}, \p{10}{b} \rightarrow \frac{163}{3145728}\Bigg]}
\end{dmath*}

\section{Discussion and conclusions}
\label{sec:concl}


Computer algebra systems are essential to many research problems in
physics and other related fields, and due to the wide variety of
technical challenges which these problems demand these tools need to
possess a large amount of flexibility and customization. One of the
major advantages of \verb+Cadabra+ is its direct connection with Python which
allows it to take advantage of the powerful constructs and the
numerous libraries which Python has. In this article we have continued
on from the discussion of the previous paper
\cite{castillo-felisola20_cadab_python_algor_gener_relat_cosmol_i},
where the applicability of \verb+Cadabra+ to calculations in general
relativity was demonstrated, to show that more general algorithms can
be constructed and exemplifying this by using tensor perturbations in
general relativity as a case study. This allowed us to show \verb+Cadabra+'s
capabilities in handling abstract tensor equations and performance
when handling very large expressions, as well as providing the
possibility of extension beyond general relativity into other fields
where perturbation theory is used such as atomic physics, condensed
matter theory and particle physics.

The central algorithm of the paper is the \verb|perturb| function from
section \ref{sec:tenspert} which provides a general way of decomposing
an object as a perturbative expansion in terms of other objects, which
in this paper has allowed us to represent many different quantities in
terms of metric perturbations. By employing the results from this we
first derived the wave-equation in the harmonic gauge for the
propagation of first order gravitational waves in the presence of
matter and their wave-relations in a vacuum. Having calculated this,
we then show that \verb+Cadabra+ is capable of handling high-order tensor
calculation by calculating coefficients for wave equations up to the
10th order.

\begin{figure}[h!]
    \centering
    \includegraphics[width=.9\textwidth]{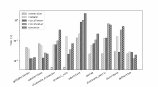} 
    \caption{Breakdown of time spent in the subroutines of \texttt{perturb}}
    \label{fig:comp_per_algo_order4}
\end{figure}

We present here some comments on the complexity of the \verb|perturb|
function and the calculation of the higher order coefficients to
illustrate the limits of how far can be explored using this technique
before the computational power required becomes excessive. We next
look at Fig. \ref{fig:comp_per_algo_order4} which shows the amount of
time spent performing various routines inside the \verb|perturb|
function for several different objects.

The non-uniformity of this graph is a result of various inherent
properties of the different expressions which mean that the various
algorithms must do different amounts of work on each object. This is
most clearly shown in the \verb|product_rule| algorithm, where the
connection which relies only on a single derivative of the metric
spends almost no time but the Riemann tensor which is composed of more
complicated derivatives of Christoffel symbols spends the most time
out of any algorithm. However, the overall behaviour is defined by the
most expensive operation which is in all cases \verb|substitute|; as
this is a property which does not depend on any property of the
expression other than its size, this shows that the dominant property
which governs the complexity of the \verb|perturb| algorithm is size
of the expression which is directly related to the number of terms in
the expression

\begin{figure}[h!]
    \centering
    \includegraphics[width=.9\textwidth]{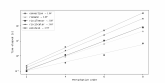} 
    \caption{Total computational time of \texttt{perturb} over a range of perturbative
      orders fitted to exponential functions}
    \label{fig:comp_pert_order_scatterplot}
\end{figure}

 We next concentrate on the graph in Fig.
 \ref{fig:comp_pert_order_scatterplot} which shows the complexity of
 the \verb|perturb| function over different perturbative orders.

As can be seen, the complexity for each object is well described by an
exponential relation with the perturbative order. This general pattern
is a consequence of the number of terms at each perturbative order. To
give a theoretical justification that this grows exponentially, we can
consider the connection given by
\begin{equation}
  \Gamma^{\mu}{}_{\nu\tau} = 
    \frac{1}{2} g^{\mu\sigma} (\partial_{\tau}{g_{\nu\sigma}} + 
      \partial_{\nu}{g_{\tau\sigma}} - 
      \partial_{\sigma}{g_{\nu\tau}})
\end{equation}
Terms in the perturbative expansion of
$\p{n}{\Gamma}^\mu{}_{\nu \tau}$ consist of a product of metric
perturbations $\p{i}{h}_{\mu \nu}$ and exactly one derivative
$\partial_{\rho}{\p{i}{h}_{\mu \nu}}$. As the sum of the individual
perturbative orders of $\p{i}{h}_{\mu \nu}$ must equal $n$, the total
number of possible terms is given by the number of ordered partitions
of $n$ which can be shown to be given by $P(n) = 2^{n-1}$. As
$\p{i}{h}_{\mu \nu}$ is symmetric there are three unique combinations
of indices for the derivative term and therefore the total number of
terms at each order is given by $3 \cdot 2^{n-1}$, which agrees very
well with the growth factor shown in the graph.

\begin{figure}[h!]
    \centering
    \includegraphics[width=.9\textwidth]{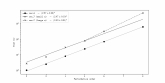} 
    \caption{Total time elapsed in the computation of the perturbed
      objects, for different orders of perturbations. The marks denote
      the computational timing, while the solid line is the fit to the
      scatter plot.}
    \label{fig:sec4_vs_sec7}
\end{figure}

Finally we make a quick comment on the behaviour shown in
Fig.~\ref{fig:sec4_vs_sec7} which shows the comparative amounts of time
it takes to run the \verb|sec4_TensorPerturbations.cnb| notebook and
the \verb|sec7_HigherOrderWaves.cnb| notebook.

The red and blue lines closely follow each other until the fifth order
where the growth rate of the time it takes for
\verb|sec7_HigherOrderWaves.cnb| to run diverges. This is a very good
illustration of the impossibility of extrapolating such graphs to
predict the runtime of higher order calculations. At the lower orders,
running \verb|sec4_TensorPerturbations.cnb| to initialise the
perturbed quantities is most expensive operation and is the dominant
behaviour which causes the two lines to be approximately parallel;
however as higher orders are probed the complexity of the computation
performed in \verb|sec7_HigherOrderWaves.cnb| which has smaller
overall constants but a higher growth factor starts to dominate
causing the divergence. It is possible that at even higher orders
there may be other parts of the calculation which scale even worse but
which are hidden at the orders we have explored.

The \verb|perturb| function is specifically intended to be as general
as possible so that it is readily customizable to suit the
requirements of fields outside of general relativity. We encourage
readers to modify the function to explore different possibilities for
the use of the algorithm. We also leave possible improvements in
efficiency and other optimisations, especially in order to minimise
the amount of memory required for large perturbative terms, to a
future work.

\subsection*{Acknowledgements}

The authors wish to thank Kasper Peeters for developing and
maintaining \verb+cadabra2+ software, and Leo Brewin, who developed
the \texttt{cdblatex.sty} \LaTeX{} package that was used extensively
in the typesetting of this work~\cite{brewin19_hybrid-latex}. MS would
like to thank E. DeLazzari, S. Vidotto, A. Quaggio and G. Casagrande
for their shared passion and teachings. The work of OCF is sponsored
by the ``Centro Científico y Tecnológico de Valparaíso'' (CCTVal),
which is funded by the grant ANID PIA/APOYO AFB180002 (Chile). This
research benefited from the grant \verb+PI_LI_19_02+ from the
Universidad Técnica Federico Santa María.


\bibliographystyle{elsarticle-num}
\bibliography{bibliography}

\begin{thebibliography}{10}
\expandafter\ifx\csname url\endcsname\relax
  \def\url#1{\texttt{#1}}\fi
\expandafter\ifx\csname urlprefix\endcsname\relax\def\urlprefix{URL }\fi
\expandafter\ifx\csname href\endcsname\relax
  \def\href#1#2{#2} \def\path#1{#1}\fi

\bibitem{castillo-felisola20_cadab_python_algor_gener_relat_cosmol_i}
O.~Castillo-Felisola, K.~Peeters, D.~T. Price, M.~Scomparin, Cadabra and python
  algorithms in general relativity and cosmology {I}: {G}eneralities (2020).

\bibitem{peeters07_cadab}
K.~Peeters, Cadabra: a field-theory motivated symbolic computer algebra system,
  Comput. Phys. Commun. 176~(8) (2007) 550.
\newblock \href {https://doi.org/10.1016/j.cpc.2007.01.003}
  {\path{doi:10.1016/j.cpc.2007.01.003}}.

\bibitem{peeters07_introd_cadab}
K.~Peeters, {Introducing Cadabra: A Symbolic Computer Algebra System for Field
  Theory problems} (2007).
\newblock \href {http://arxiv.org/abs/hep-th/0701238}
  {\path{arXiv:hep-th/0701238}}.

\bibitem{peeters07_symbol_field_theor_with_cadab}
K.~Peeters,
  \href{http://www.fachgruppe-computeralgebra.de/CA-Rundbrief/car41.pdf}{Symbolic
  field theory with cadabra}, Computeralgebra Rundbrief 41 (2007) 16.
\newline\urlprefix\url{http://www.fachgruppe-computeralgebra.de/CA-Rundbrief/car41.pdf}

\bibitem{Brewin:2019qbs}
L.~Brewin, {Using Cadabra for tensor computations in General Relativity}
  (2019).
\newblock \href {http://arxiv.org/abs/1912.08839} {\path{arXiv:1912.08839}}.

\bibitem{Abbott:2016blz}
B.~P. Abbott, et~al., {Observation of Gravitational Waves from a Binary Black
  Hole Merger}, Phys. Rev. Lett. 116~(6) (2016) 061102.
\newblock \href {http://arxiv.org/abs/1602.03837} {\path{arXiv:1602.03837}},
  \href {https://doi.org/10.1103/PhysRevLett.116.061102}
  {\path{doi:10.1103/PhysRevLett.116.061102}}.

\bibitem{Abbott:2016nmj}
B.~P. Abbott, et~al., {GW151226: Observation of Gravitational Waves from a
  22-Solar-Mass Binary Black Hole Coalescence}, Phys. Rev. Lett. 116~(24)
  (2016) 241103.
\newblock \href {http://arxiv.org/abs/1606.04855} {\path{arXiv:1606.04855}},
  \href {https://doi.org/10.1103/PhysRevLett.116.241103}
  {\path{doi:10.1103/PhysRevLett.116.241103}}.

\bibitem{Abbott:2017vtc}
B.~P. Abbott, et~al., {GW170104: Observation of a 50-Solar-Mass Binary Black
  Hole Coalescence at Redshift 0.2}, Phys. Rev. Lett. 118~(22) (2017) 221101,
  [Erratum: Phys. Rev. Lett.121,no.12,129901(2018)].
\newblock \href {http://arxiv.org/abs/1706.01812} {\path{arXiv:1706.01812}},
  \href {https://doi.org/10.1103/PhysRevLett.118.221101,
  10.1103/PhysRevLett.121.129901} {\path{doi:10.1103/PhysRevLett.118.221101,
  10.1103/PhysRevLett.121.129901}}.

\bibitem{TheLIGOScientific:2017qsa}
B.~P. Abbott, et~al., {GW170817: Observation of Gravitational Waves from a
  Binary Neutron Star Inspiral}, Phys. Rev. Lett. 119~(16) (2017) 161101.
\newblock \href {http://arxiv.org/abs/1710.05832} {\path{arXiv:1710.05832}},
  \href {https://doi.org/10.1103/PhysRevLett.119.161101}
  {\path{doi:10.1103/PhysRevLett.119.161101}}.

\bibitem{GBM:2017lvd}
B.~P. Abbott, et~al., {Multi-messenger Observations of a Binary Neutron Star
  Merger}, Astrophys. J. 848~(2) (2017) L12.
\newblock \href {http://arxiv.org/abs/1710.05833} {\path{arXiv:1710.05833}},
  \href {https://doi.org/10.3847/2041-8213/aa91c9}
  {\path{doi:10.3847/2041-8213/aa91c9}}.

\bibitem{Monitor:2017mdv}
B.~P. Abbott, et~al., {Gravitational Waves and Gamma-rays from a Binary Neutron
  Star Merger: GW170817 and GRB 170817A}, Astrophys. J. 848~(2) (2017) L13.
\newblock \href {http://arxiv.org/abs/1710.05834} {\path{arXiv:1710.05834}},
  \href {https://doi.org/10.3847/2041-8213/aa920c}
  {\path{doi:10.3847/2041-8213/aa920c}}.

\bibitem{Abbott:2020khf}
R.~Abbott, et~al., {GW190814: Gravitational Waves from the Coalescence of a 23
  Solar Mass Black Hole with a 2.6 Solar Mass Compact Object}, Astrophys. J.
  Lett. 896~(2) (2020) L44.
\newblock \href {http://arxiv.org/abs/2006.12611} {\path{arXiv:2006.12611}},
  \href {https://doi.org/10.3847/2041-8213/ab960f}
  {\path{doi:10.3847/2041-8213/ab960f}}.

\bibitem{Arcos:2015uqa}
H.~Arcos, M.~Kr\v{s}\v{s}\'ak, J.~G. Pereira, {Exploring Higher-Order
  Gravitational Waves} (2015).
\newblock \href {http://arxiv.org/abs/1504.07817} {\path{arXiv:1504.07817}}.

\bibitem{THOULESS1960553}
D.~J. Thouless,
  \href{https://www.sciencedirect.com/science/article/pii/0003491660901226}{Perturbation
  theory in statistical mechanics and the theory of superconductivity}, Annals
  of Physics 10~(4) (1960) 553--588.
\newblock \href {https://doi.org/https://doi.org/10.1016/0003-4916(60)90122-6}
  {\path{doi:https://doi.org/10.1016/0003-4916(60)90122-6}}.
\newline\urlprefix\url{https://www.sciencedirect.com/science/article/pii/0003491660901226}

\bibitem{BOX200295}
M.~A. Box, Radiative perturbation theory: a review, Environmental Modelling
  Software 17~(1) (2002) 95--106.
\newblock \href {https://doi.org/https://doi.org/10.1016/S1364-8152(01)00056-1}
  {\path{doi:https://doi.org/10.1016/S1364-8152(01)00056-1}}.

\bibitem{Picasso2009}
L.~E. Picasso, L.~Bracci, E.~d'Emilio,
  \href{https://doi.org/10.1007/978-0-387-30440-3_402}{Perturbation Theory in
  Quantum Mechanics}, Springer New York, New York, NY, 2009, pp. 6723--6747.
\newblock \href {https://doi.org/10.1007/978-0-387-30440-3_402}
  {\path{doi:10.1007/978-0-387-30440-3_402}}.
\newline\urlprefix\url{https://doi.org/10.1007/978-0-387-30440-3_402}

\bibitem{Cvetic:2011vz}
M.~Cvetic, J.~Halverson, {TASI Lectures: Particle Physics from Perturbative and
  Non-perturbative Effects in D-braneworlds}, in: {Theoretical Advanced Study
  Institute in Elementary Particle Physics}: {String theory and its
  Applications: From meV to the Planck Scale}, 2011.
\newblock \href {http://arxiv.org/abs/1101.2907} {\path{arXiv:1101.2907}},
  \href {https://doi.org/10.1142/9789814350525_0005}
  {\path{doi:10.1142/9789814350525_0005}}.

\bibitem{1975STIA}
M.~{van Dyke}, {Perturbation methods in fluid mechanics /Annotated edition/},
  NASA STI/Recon Technical Report A 75 (1975) 46926.

\bibitem{Ayers}
P.~W. Ayers, J.~S.~M. Anderson, L.~J. Bartolotti, Perturbative perspectives on
  the chemical reaction prediction problem, International Journal of Quantum
  Chemistry 101~(5) (2005) 520--534.
\newblock \href {https://doi.org/https://doi.org/10.1002/qua.20307}
  {\path{doi:https://doi.org/10.1002/qua.20307}}.

\bibitem{Weinberg:1972kfs}
S.~Weinberg, {Gravitation and Cosmology}, John Wiley and Sons, New York, 1972.

\bibitem{Carroll:2004st}
S.~M. Carroll, {Spacetime and Geometry}, Cambridge University Press, 2019.

\bibitem{brewin19_hybrid-latex}
L.~Brewin, hybrid-latex, \url{https://github.com/leo-brewin/hybrid-latex}
  (2019).

\end{thebibliography}





\end{document}